\documentclass[twocolumn]{aastex63}
\pdfoutput=1
\usepackage{amssymb,amsmath}
\usepackage{bm}

\shorttitle{Propagation effects in the FRB 20121102A spectra}
\shortauthors{Levkov et al.}

\begin{document}

\reportnum{INR-TH-2020-040}
\reportnum{arXiv:2010.15145}

\title{Propagation effects in the FRB 20121102A spectra}

\correspondingauthor{D.G. Levkov}\email{levkov@ms2.inr.ac.ru}

\author{D.G. Levkov}
\affiliation{Institute for Nuclear Research of the Russian Academy
  of Sciences, Moscow 117312, Russia}
\affiliation{Institute for Theoretical and Mathematical Physics, MSU,
  Moscow 119991, Russia} 
\author{A.G. Panin}       
\affiliation{Institute for Nuclear Research of the Russian Academy
  of Sciences, Moscow 117312, Russia}
\affiliation{Moscow Institute of Physics and Technology, 
  Dolgoprudny 141700, Russia}
\author{I.I. Tkachev}       
\affiliation{Institute for Nuclear Research of the Russian Academy
  of Sciences, Moscow 117312, Russia}
\affiliation{Novosibirsk State University, Novosibirsk 630090, Russia}

\begin{abstract}
We advance theoretical methods for studying propagation effects in the
Fast Radio Burst (FRB) spectra. We derive their autocorrelation
function in the model with diffractive lensing and strong
Kolmogorov--type scintillations and analytically obtain the spectra
lensed on different plasma density profiles. With these tools, we
reanalyze the highest frequency 4--8 GHz data of~\cite{Gajjar:2018bth}
for the repeating FRB~20121102A (FRB 121102). In the data we discover,
first, a remarkable spectral structure of almost equidistant peaks
separated by ${95\pm 16}$~MHz. We suggest that it can originate from
diffractive lensing of the FRB signals on a compact gravitating object
of mass $10^{-4}\, M_\odot$ or on a plasma underdensity near the
source. Second, the spectra include erratic interstellar, presumably
Milky Way scintillations. We extract their decorrelation bandwidth
${3.3\pm 0.6}$~MHz at reference frequency 6~GHz. The third feature is
a GHz--scale pattern which, as we find, linearly drifts with time and
presumably represents a wide--band propagation effect,
e.g.\ GHz--scale scintillations. Fourth, many spectra are dominated by
a narrow peak at 7.1~GHz. We suggest that it can be caused by a
propagation through a plasma lens, e.g.,\ in the host galaxy. Fifth,
separating the propagation effects, we give strong arguments that the
intrinsic progenitor spectrum has  narrow GHz bandwidth and variable
central frequency. This confirms expectations from the previous
observations. We discuss alternative interpretations of the above
spectral features.
\bigskip
\bigskip
\end{abstract}


\section{Introduction}

\label{sec:introduction}

Fast Radio Bursts (FRB) still mystify researchers due to 
unknown nature of their progenitors and anticipation for new
propagation effects that may enrich their signals with information
on the traversed medium, see reviews by~\cite{Popov:2018hkz},
  \cite{Petroff:2019tty}, and \cite{Cordes:2019cmq}. Sky
  distribution of the registered  bursts is
  isotropic \citep{Thornton:2013iua, Shannon:2018} implying that they
  travel cosmological (Gpc) distances, and localization of  several FRB 
sources confirms that, see~\cite{Chatterjee:2017dqg, Marcote:2017wan,  
  Tendulkar:2017vuq, Bannister:2019iju, Ravi:2019alc, Prochaska:2019,
  Marcote:2020ljw, Bhandari:2020oyb}. As a consequence,
the FRB propagation effects include multi--scale
scintillations~\citep{Rickett1990, Narayan,  Lorimer-Kramer, Woan}
i.e.\ diffractive scattering of radio waves on the turbulent plasma
clouds in the FRB host galaxy, Milky Way, and in the intervening galactic
halos. In addition, the FRB
waves may be lensed by refractive plasma 
clouds with smooth profiles~\citep{Clegg:1997ya, 
  Cordes:2017eug}, or lensed gravitationally by exotic massive
compact objects like primordial black holes or dense  
mini-halos~\citep{Zheng:2014rpa, Munoz:2016tmg, Eichler:2017eid,
  Katz:2019qug}. Thus, studying the FRB signals one may hope to learn
something about the cosmological  parameters~\citep{Deng:2013aga, Yu:2017beg, 
  Walters:2017afr, Macquart:2020lln}, 
intergalactic medium~\citep{Zhou:2014yta, Zheng:2014rpa,
  Akahori:2016ami, Fujita:2016yve}, exotic inhabitants of the
intergalactic space~\citep{Zheng:2014rpa, Munoz:2016tmg,
  Eichler:2017eid, Katz:2019qug}, and plasma in the far-away
galaxies. 

The scales of the FRB events point at extreme conditions in their
progenitors~\citep{Platts:2018hiy} which are hard to achieve in
realistic astrophysical settings, see~\cite{Ghisellini:2017twg},
  \cite{Lu:2017prv}, \cite{Katz:2018afd} and cf.~\cite{Yang:2017tmb},
  \cite{Wang:2019bpi}. Millisecond durations of the
  bursts~\citep{Cho:2020gtg} constrain the progenitor sizes to be 100
  km or less in the absence of special relativistic
    effects. Besides, FRB microsecond substructure observed by
    \cite{Nimmo:2020sva} limits the instantaneous emission 
    regions down to 1 km. High  
spectral fluxes (${\sim \mbox{Jy}}$) at GHz frequencies give record
brightness temperatures $T\sim 10^{35}-10^{41}\,
\mbox{K}$~\citep{Cordes:2019cmq, Nimmo:2021yob} and therefore
support non--thermal, presumably coherent emission mechanisms. The fluxes imply
strong~\citep{Yang:2020uqt} electromagnetic fields $10^{13} \,
\mbox{V}/\mbox{m}$ near the sources which nevertheless should not halt the
emission~\citep{Ghisellini:2017twg}. In addition, the periodic activity of
the two repeating FRB sources~\citep{Amiri:2020gno, Cruces:2020gmn}
points at rotational motions of compact objects and further
  restricts the progenitor models. Recently an unusually
  intense radio burst was observed from a Galactic magnetar
  SGR~1935+2154 \citep{CHIMEFRB:2020abu, Bochenek:2020zxn,
    Kirsten:2020yin} thus marking these objects as main 
  candidates for the FRB sources. Urge to explain the FRB properties
  and absence of generally accepted theoretical models requires new
  data. And this field progresses rapidly, see
  e.g.\ \cite{Tendulkar:2020npy}, \cite{Pleunis:2020vug},
  \cite{Kirsten:2021llv}, and \cite{Rafiei-Ravandi:2021hbw}.

Critical information on the FRB central engines can be delivered
by their spectra, although a careful data analysis is needed to
  separate the propagation effects. So far, spectral properties of
FRB received undeservingly little attention in the literature, where 
promising results started to appear only recently,
see~\cite{Gajjar:2018bth}, \cite{Hessels:2018mvq},
\cite{Chawla:2020rds}, \cite{Majid:2020dwq},  \cite{Pearlman:2020sox},
and \cite{Pleunis:2020vug}.

With this paper, we develop methods for studying FRB spectral
structures and spectral propagation effects, see
also~\cite{Katz:2019qug}. We apply these tools to investigate the
frequency spectra of the repeating FRB 20121102A commonly known as
FRB~121102~\citep{Spitler:2016dmz, Chatterjee:2017dqg}. The bursts
were registered at  $4 - 8$~GHz by the Breakthrough Listen Digital
Backend at the Green Bank  Telescope by~\cite{Gajjar:2018bth}.

The paper is organized as follows. We introduce the FRB spectra in
Sec.~\ref{sec:spectra-frb-20121102A} and consider their dominating 7.1~GHz
peak in Sec.~\ref{sec:main-peak}. Wide--band
pattern and the progenitor spectrum are considered in
Sec.~\ref{sec:progenitor-spectrum}. In Sec.~\ref{sec:scintillations}
we study narrow--band scintillations. New periodic spectral structure
is analyzed in Sec.~\ref{sec:periodic-structure}. In
Sec.~\ref{sec:comp-with-earl} we compare our narrow--band spectral  
  analysis with the other results. We summarize in
Sec.~\ref{sec:conclusions}. Appendices describe theoretical models
used to fit the experimental data.

\section{Spectra of  FRB 20121102A} 
\label{sec:spectra-frb-20121102A}
\begin{figure}[t]
  \centerline{\includegraphics{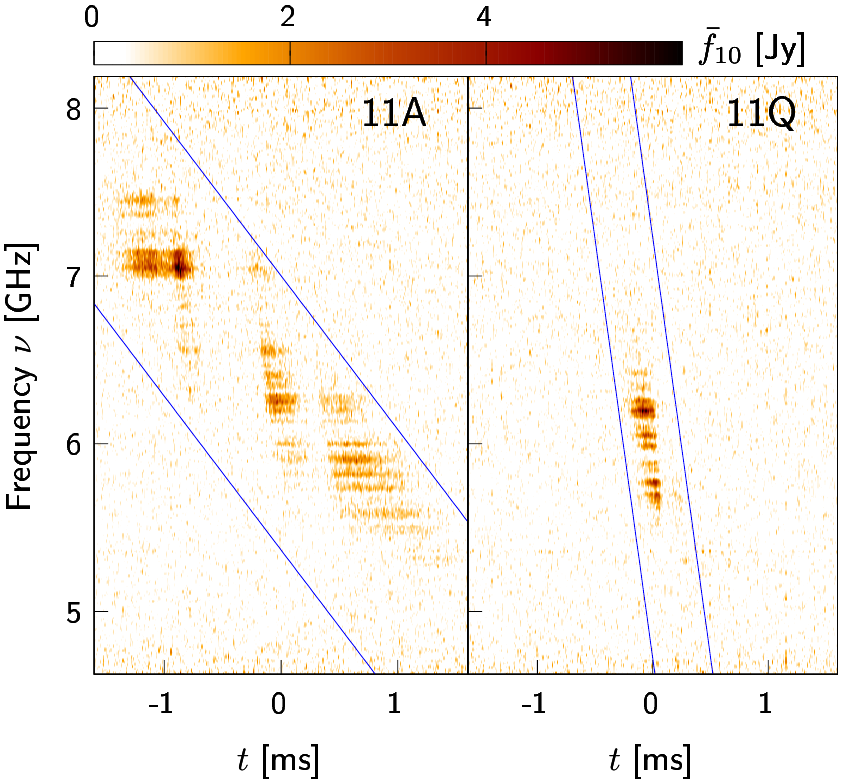}}
  \caption{Spectral flux densities $\bar{f}_{10}(t,\, \nu)$ of the bursts
    11A and~11Q. 
    \label{fig:spectra_2d}}
\end{figure}

Public data of the \cite{data}     
give de--dispersed spectral flux density\footnote{Measured in Jansky $\mbox{Jy} = 
  10^{-26}\,    \mbox{W}/(\mbox{m}^2\, \mbox{Hz})$.} $f(t, \, \nu)$ of
FRB 20121102A
as a function of time $t$ and frequency $\nu$. The provided time
intervals include 18 bursts\footnote{Later \cite{Zhang:2018jux}
    published another 72~bursts from the same observing session.
    But these are too weak for the spectral analysis performed in
    this paper.}
within the first 60 minutes of the 6--hour 
observations on August 26, 2017. The bursts are assigned
identifiers\footnote{The  bursts 11L, 11P, and 11R are absent in the
  public data.} 11A through 11R and 12A through 12C, in order of their
arrival. In the wide--band analysis, we suppress the instrumental
  noise using Gaussian average over the moving frequency
window $\sigma =10\, \mbox{MHz}$,
\begin{equation}
  \label{eq:2}
  \bar{f}_\sigma(t,\, \nu) = \int d\nu' \, f(t,\, \nu+\nu') \;
  \frac{\mathrm{e}^{-\nu'^2 / 2\sigma^2}}{\sigma\sqrt{2\pi}}\;.
\end{equation}
By construction, $\bar{f}_{10}$ fairly represents $f$ on scales
exceeding $\Delta \nu \geq \pi \sigma \sqrt{2} \approx 45\,
\mbox{MHz}$. The modulations at smaller scales $\Delta
\nu$ are suppressed by a factor $\exp(-2\pi^2 \sigma^2 / \Delta
\nu^2)$. We stress that this smoothing is not used in the analysis
  of narrow--band scintillations. Note that  unlike the simplest
binning, Eq.~(\ref{eq:2}) does not rely on 
a preselected grid of frequencies and therefore does not create bias in
the discussion of the spectral periodicity\footnote{It is worth noting that
  our smooth spectra are practically identical to the ones produced
  by $\simeq 23 \, \mbox{MHz}$ binning, or by Savitzky--Goley filtering
  with appropriate parameters.}. The color--coded smooth flux
densities $\bar{f}_{10}(t,\, \nu)$ of the bursts 11A and 11Q are
shown\footnote{In what follows we mostly ignore complex temporal
    structure of the burst spectra in Fig.~\ref{fig:spectra_2d}a. In
    particular, many events include sub--bursts appearing at lower
    frequencies at later times (the ``sad trombone''
    effect, see~\cite{Hessels:2018mvq} and \cite{Josephy:2019ahz}).} in
Fig.~\ref{fig:spectra_2d}.   

\begin{figure}[b!]
  \includegraphics{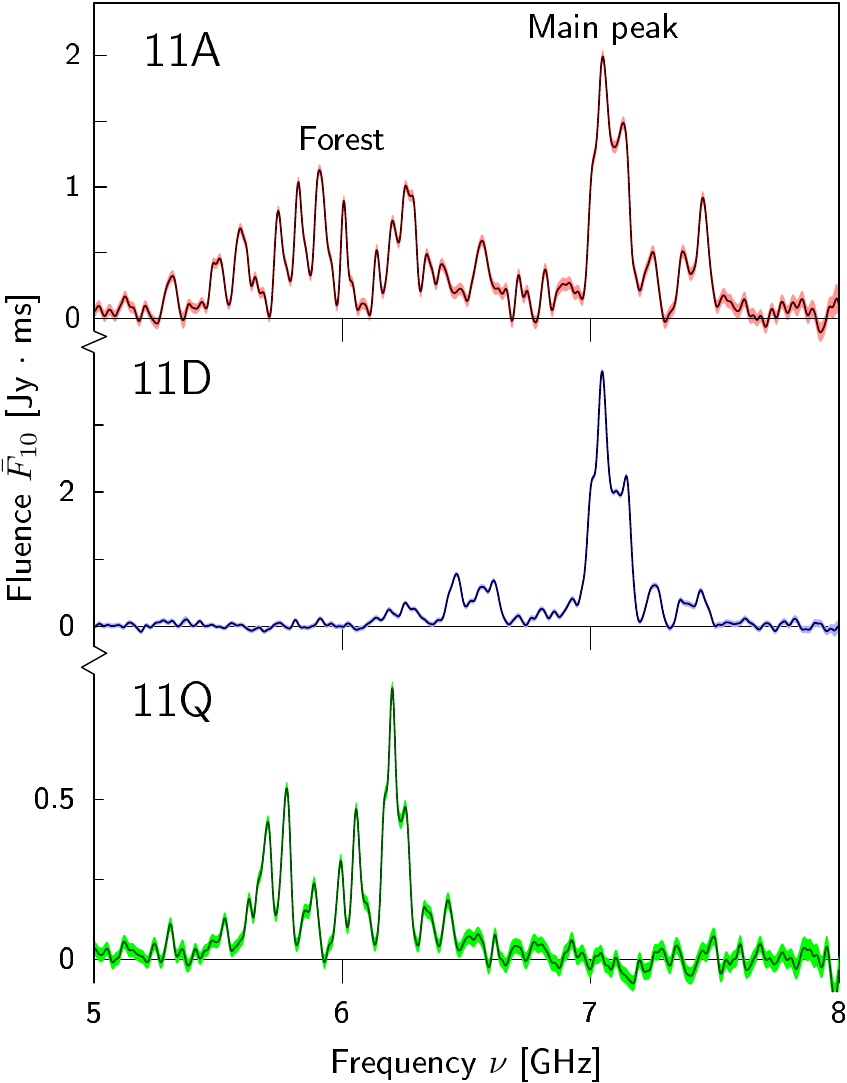}
\caption{Smoothed spectra of the bursts 11A, 11D, and 11Q. 
  Instrumental errors are displayed with shaded areas near the
  graphs. Note different scales on the vertical axes.}
 \label{fig:spectrumAC}
\end{figure}

To further visualize the FRB spectra, we integrate $f$ over the 
burst duration $t_1 < t < t_2$ and obtain its spectral fluence
$F(\nu)$ i.e.\ the burst total energy per unit frequency,  
\begin{equation}
  \label{eq:1}
  \bar{F}_\sigma(\nu) = \int\limits_{t_1(\nu)}^{t_2(\nu)} dt \, \bar{f}_\sigma(t,\,
  \nu)\;,
\end{equation}
where the bar again denotes the smoothing Eq.~(\ref{eq:2}). The signal
region ${t_1(\nu) < t <   t_2(\nu)}$ (tilted lines in
Fig.~\ref{fig:spectra_2d}) is chosen in
Appendix~\ref{sec:computing-spectra} to minimize the background 
noise. The size of this region is about a millisecond, varying from
burst to  burst. Outside of the signal region $\bar{f}_{10}$
fluctuates around zero. The spectral fluences $\bar{F}_{10}$ of
the bursts 11A, 11D,  and 11Q are demonstrated in
Fig.~\ref{fig:spectrumAC}, where the shaded areas near the graphs
represent instrumental errors. Apparently, the latter are small and we
will ignore  them in what follows.

The spectra in Fig.~\ref{fig:spectrumAC} expose a number of unusual
features. First, the graphs 11A and 11D include high and narrow
``main peak" at $\nu \approx 7.1\; \mbox{GHz}$. In fact, 10 out of
the 18 spectra have this feature precisely at the same position. In  some events, 
e.g.\ 11D or 11F, this peak carries most of the burst energy. On the
other hand, the remaining 8 bursts have no 7.1~GHz peak at all,
see the graph 11Q in Fig.~\ref{fig:spectrumAC}. Second, almost 
all spectra display ``forests'' of smaller peaks of width $\lesssim 100$~MHz.
The forests  have physical origin, since their presence is not
sensitive to the smoothing window $\sigma$ in
Eq.~(\ref{eq:2}). Third, the envelopes of the forests have distinctive
near--parabolic forms with cutoffs at low and high frequencies. Below
we study and explain these three properties. 

\begin{figure}
  \centerline{\includegraphics{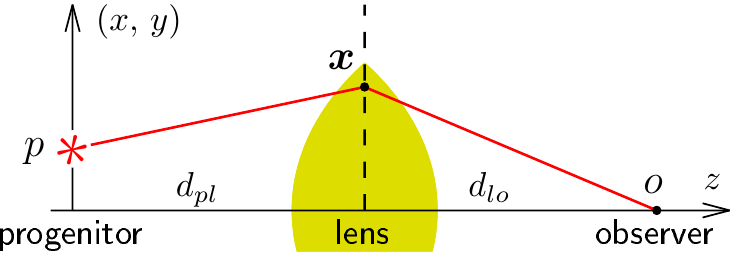}}
  \caption{Plasma lens bending the radio ray
    $p\bm{x}o$. \label{fig:lens}}
\end{figure}

\section{Main peak from the plasma lens}
\label{sec:main-peak}
The dominating feature of the most spectra is a high and narrow
``main'' peak at 7.1~GHz, see the top two panels in
Fig.~\ref{fig:spectrumAC}. The same peak was observed previously by
  \cite{Gajjar:2018bth}, but it was never explained. Let us argue that
it may result from a propagation of the FRB signal through a
plasma lens~\citep{Clegg:1997ya, Cordes:2017eug}. Indeed, the latter
usually splits the radio wave into multiple rays. Even if the
interference of the rays is not relevant, their coalescence at
certain frequencies~--- the lens caustics~--- may produce high
spectral spikes of a specific form.

We consider the lens of~\cite{Clegg:1997ya} and~\cite{Cordes:2017eug}
with dispersion measure depending on one transverse coordinate $x$:
${\mathrm{DM}(x) = \mathrm{DM}_l\, \mathrm{e}^{-x^2/a^2}}$, see
Fig.~\ref{fig:lens}. It has two parameters: the size $a$ and central 
dispersion $\mathrm{DM}_l$. Such one--dimensional lenses are often
  used for modeling plasma overdensities, cf.~\cite{Cordes:2017eug}.
Note that occulting AU--sized structures are expected to be present
  in turbulent galactic plasmas, and one of them may get on the way of
  FRB~20121102A. In fact, lensing on such structures is consistent with
  extreme scattering events observed in the light curves of some
  active galactic nuclei~\citep{1987Natur.326..675F,
    Bannister:2016zqe} and perturbations in pulsar
  timings~\citep{Coles:2015uia}. Alternatively, the lens may represent
  long ionized filament in the supernova remnant from the host galaxy, 
  cf.~\cite{2011MNRAS.410..499G} and~\cite{Michilli:2018zec}.
Generically, the lens is located in the FRB host galaxy
or in the Milky Way, at distances $d_{pl}$ from  the source and
$d_{lo}$ from us; $d_{po} = d_{pl} + d_{lo}$. 

The radio wave receives a dispersive time delay in the lens and as a
consequence, propagates along the bended path $p\bm{x}o$ in
Fig.~\ref{fig:lens}.  These two effects give the phase shift,
see~\cite{Clegg:1997ya}, \cite{Cordes:2017eug}, and
Appendix~\ref{sec:lenses} for details, 
\begin{equation}
  \label{eq:3}
  \Phi_l(\bm{x}) = \frac{1}{2r_{F,\, l}^2} \left[
    - \alpha_l a^2\, \mathrm{e}^{-x^2/a^2} + (\bm{x} - \tilde{\bm{x}})^2\right]\;,
\end{equation}
where the second term comes from the ray geometry, ${\bm{x}   =(x,\,
  y)}$ is the transverse coordinate in the lens plane and its value 
\begin{equation}
  \label{eq:16}
  \tilde{\bm{x}} = (\bm{x}_p d_{lo} + \bm{x}_o d_{pl})/d_{po}\;. 
\end{equation}
corresponds to a straight propagation between the transverse positions $\bm{x}_p$
and $\bm{x}_o$ of the progenitor and the observer.
Note that $\tilde{\bm x}$ coincides with $\bm{x}_p$ when the lens
resides in the FRB host galaxy, and $\tilde{\bm x} \approx   
{\bm x}_o$ if it is close to us. We also introduced the 
lens Fresnel scale $r_{F,\, l} = (d_{pl} d_{lo}/2\pi \nu
d_{po})^{1/2}$ and a dimensionless parameter   
\begin{equation}
  \label{eq:4}
  \alpha_l = \frac{2 e^2 r_{F,\, l}^2}{a^2 m_e \nu} \; \mathrm{DM}_l
\end{equation}
characterizing the lens dispersion. Note that $\alpha_l^{-1/2} \propto \nu$ is a
dimensionless analog of frequency. 

In ~\cite{Clegg:1997ya} and \cite{Cordes:2017eug} the 
lens Eq.~(\ref{eq:3}) was solved in the limit of geometric optics  $a \gg
r_{F,\,   l}$. We shortly describe this solution below and give a
detailed review in Appendix~\ref{sec:lenses}. In the geometric limit the
radio  waves go along the definite paths $\bm{x}_j = (x_j ,\, y_j)$
extremizing the phase Eq.~(\ref{eq:3}). Since the one--dimensional lens bends
the rays only in the $x$ direction, $y_j = \tilde{y}$ corresponds to a
straight propagation, and $x_j$ satisfies the nonlinear equation
$\partial_x\Phi_l = 0$. One may solve the latter graphically, by
plotting $\Phi_l(x)$ and identifying the extrema.

Within this approach one can explicitly see that as long as the shift
$\tilde{x}$ is small, there exists only one extremal radio path ${x =
  x_1}$ at any~$\alpha_l$. But above the critical shift $ \tilde{x} > 
{\tilde{x}_{cr} = a(3/2)^{3/2}}$ another two solutions $x_2,\, x_3$   
appear at ${\alpha_-(\tilde{x}) < \alpha_l < \alpha_+(\tilde{x})}$
i.e.\ inside a certain frequency interval. Thus, the radio waves with
these frequencies propagate along three different paths. The two
additional paths coincide, ${x_2    = x_3}$, at the interval
boundaries $\alpha_{\pm}$. Besides, the three--path frequency interval is
vanishingly small ($\alpha_+ = \alpha_-$) at  $\tilde{x} =
\tilde{x}_{cr}$ but becomes  larger in size at larger shifts
$\tilde{x}$.   

\begin{figure}
  \centerline{\includegraphics[width=8.6cm]{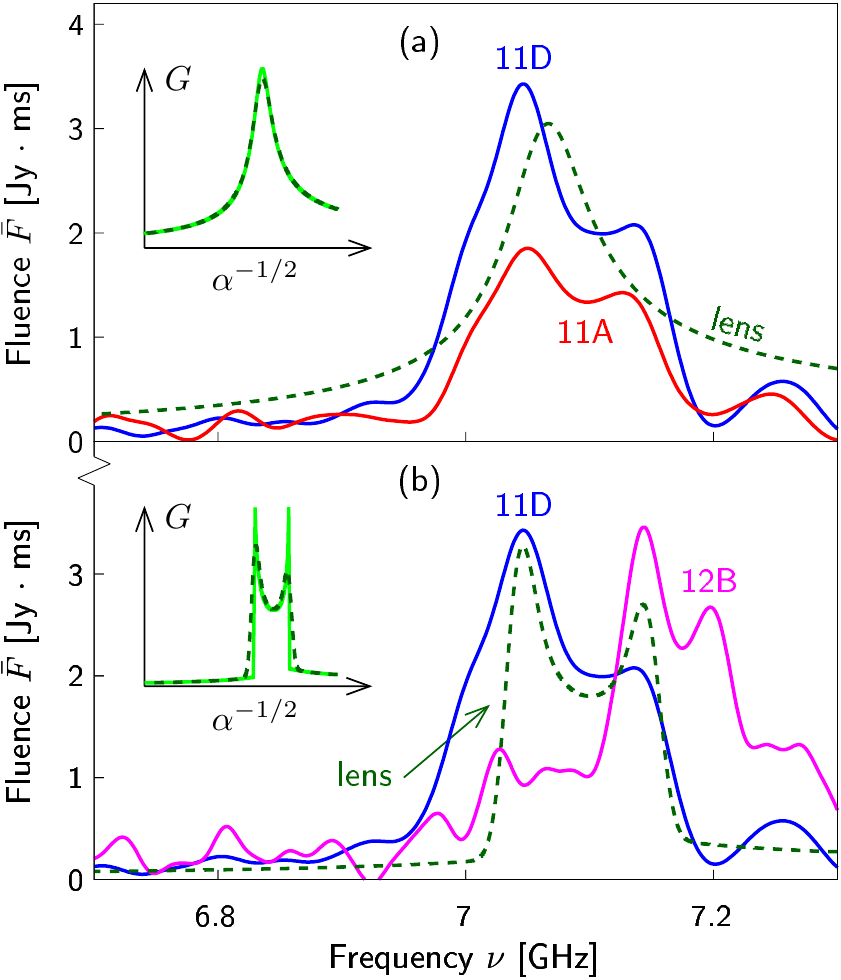}}
  \caption{Main peaks of the bursts 11A, 11D, and 12B (solid
    lines). The graphs 11D are fitted with the smoothed theoretical
    spectra $\bar{G}_{10}(\nu) F_p$ of the lens  (dashed) at
    (a)~${\tilde{x} < \tilde{x}_{cr}}$ and (b)~$\tilde{x} >
    \tilde{x}_{cr}$, where $F_p$ is a constant. The insets show
    $G$~(solid) and $\bar{G}_{10}$ (dashed) as functions of 
    $\alpha_l^{-1/2}$ in the respective cases. \label{fig:main_peak}}
\end{figure}

From the observational viewpoint, the lens focuses or disperses the
FRB signal along each path multiplying the intrinsic progenitor fluence
$F_{p}$ with the gain factor: ${F = G(\nu) F_p}$. In the
refractive one--dimensional case
the theoretical gain factor 
\begin{equation}
  \label{eq:11}
  G(\nu) = r_{F,\, l}^{-2}\sum_j |\partial_x^2\Phi_{l}(x_j)|^{-1}
\end{equation}
involves second derivatives of the phase at~$x=x_j$, where we ignore
the interference. Thus, the function $G(\nu)$ is infinite at
the lens caustics $\partial_x^2 \Phi_l = \partial_x \Phi_l = 0$ where two
radio paths~--- the extrema $x_j$ of the phase~--- coalesce. This
regime takes place at $\tilde{x} > \tilde{x}_{cr}$ when $G$ becomes
infinite at the   ``frequencies'' $\alpha_\pm^{-1/2}$ due to
coalescence of the paths  $x_2$ and $x_3$. The respective graph of $G$
has a particular two--spike form shown in the inset of
Fig.~\ref{fig:main_peak}b. In reality, the singularities of $G(\nu)$
are regulated by the instrumental resolution / smoothing in
Eq.~(\ref{eq:2}) (dashed line in the figure) and wave effects. The
regime $\tilde{x} < \tilde{x}_{cr}$ is entirely different,
  however. In this case the function $G(\nu)$ is smooth, see the
inset in Fig.~\ref{fig:main_peak}a. 

The shape of the main peak in the experimental data looks similar to
$G(\nu)$. It is particularly tempting to identify the side spikes
$\nu_0 \pm \Delta \nu/2$ of this peak with the positions of the lens
caustics  in the regime ${\tilde{x} > \tilde{x}_{cr}}$. The maxima  
of the graph 11D in Fig.~\ref{fig:main_peak}b give ${\nu_0 = 7.095\;
  \mbox{GHz}}$ and ${\Delta \nu/\nu_0 = 0.0137}$. In
Appendix~\ref{sec:lens-caustics} we derive analytic expressions for
the caustic positions at ${\Delta \nu \ll \nu_0}$: Eqs.~\eqref{eq:6} and
\eqref{eq:12}. With the above experimental numbers, they give the
source (observer) shift ${\tilde{x}/a  \approx 1.885}$ and a combination 
of the lens parameters  
\begin{equation}
  \label{eq:7}
    \beta \equiv \left(\frac{\mbox{DM}_l}{\mbox{pc} \cdot \mbox{cm}^{-3}} \right)
  \left(\frac{a}{\mbox{AU}}\right)^{-2}
  \left(\frac{\min\{d_{pl}, \, d_{lo}\}}{\mbox{kpc}} \right)
\end{equation}
entering $\alpha_l$ in Eq.~(\ref{eq:4}): $\beta \approx
0.0355$. Notably, the latter value is consistent with the parameters
of the AU--sized structures explaining the extreme scattering
  events~\citep{1987Natur.326..675F, Bannister:2016zqe,
    Coles:2015uia} and parameters of the supernova
  filaments, cf.~\cite{2011MNRAS.410..499G}.

There is another, qualitatively different fit of the main peak with
the lens spectrum. Namely, if $\tilde{x}$ is slightly below critical,
the function $G(\nu)$ has a narrow maximum $\nu = \nu_0$ with
half--height width $\Delta\nu' \ll \nu$ near the point where the caustics
are about to appear, see the inset in Fig.~\ref{fig:main_peak}a. One
can therefore interpret  the major part of the 7.1~GHz peak as
the effect of the lens with ${\tilde{x} < \tilde{x}_{cr}}$, ignoring the
side spikes. In Sec.~\ref{sec:periodic-structure} we will see that 
the latter spikes correlate with the short--scale periodic structure of
the spectra, so they may be unrelated to a refractive lensing, indeed. We read
off $\nu_0 \approx 7.066\, \mbox{GHz}$ and $\Delta 
\nu'/\nu_0 \approx 0.014$ from the spectrum 11D in
Fig.~\ref{fig:main_peak}a and use Eqs.~(\ref{eq:13}), (\ref{eq:14}) of
Appendix~\ref{sec:lenses} to compute the lens parameters in this
regime: $\beta \approx 0.031$ and $\tilde{x}/a \approx 1.81$. 

It is worth recalling that the experimental spectra in
Fig.~\ref{fig:main_peak} involve smoothing over the frequency intervals
${\sigma=10\, \mbox{MHz}}$. To perform the precise comparison, we 
smooth the theoretical lens spectra $G(\nu) F_p$ in the same way and
then fit them to the graph 11D, see the dashed lines in
Fig.~\ref{fig:main_peak}. This procedure almost does not affect the
one--peak fit
in Fig.~\ref{fig:main_peak}a but essentially modifies the
caustics in the double--peaked lens spectrum in
Fig.~\ref{fig:main_peak}b. We obtain an improved estimate of the lens
parameters in the latter case: $\tilde{x}/a \approx 1.890$ and $\beta
\approx 0.0359$. In what follows we determine $\tilde{x}/a$ and
$\beta$ using 
the smoothed double--peaked lens spectrum.

So far we completely disregarded the interference of the lensed radio rays
which may lead to oscillations of the gain factor with frequency. Note,
however, that multiple rays exist only at $\tilde{x}> \tilde{x}_{cr}$
in the narrow frequency interval between the caustics, and we do not
see any oscillatory behavior there. To no surprise, since the
smoothing Eq.~(\ref{eq:2}) destroys any oscillations on scales below 
$45 \,\mbox{MHz}$. Requiring the interference period to be smaller, we
obtain a constraint $a/r_{F,\, l} \gtrsim (\nu/\sigma)^{1/2} \sim 25$
or~${(a/\mbox{AU})^2 \gtrsim 0.007 \cdot \min(d_{pl},\,
  d_{lo})/\mbox{kpc}}$.

\begin{figure}
  \centerline{\includegraphics[width=8.6cm]{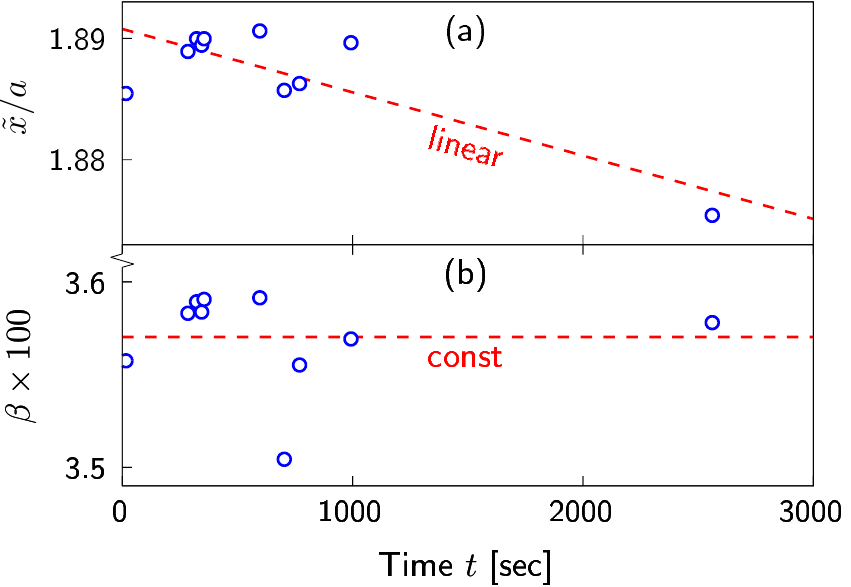}}
  \caption{Parameters $\tilde{x}/a$ and $\beta$ of the lens
    extracted from the main peaks of different burst spectra. Here
      we use the double--peaked fits at ${\tilde{x} >
        \tilde{x}_{cr}}$. Note that the main peak is absent 
        in the large interval between the bursts 11M and~12B. We will
        explain this property in Sec.~\ref{sec:progenitor-spectrum}.}
    \label{fig:7GHz}
\end{figure}
To test the lens hypothesis, we compare the main peaks
in different FRB spectra. Generically, one expects to find
almost time--independent $\beta$ and linearly evolving $\tilde{x}/a$
due to  transverse motion of the source/observer with respect to the
lens. In Fig.~\ref{fig:7GHz} we plot these parameters 
extracted from the double--peaked fits (Fig.~\ref{fig:main_peak}b) of
different  spectra. All values of $\beta$ and $\tilde{x}/a$ 
are almost identical except for the burst 12B, see the rightmost
points in Figs.~\ref{fig:7GHz}a,b. The ``main'' peak in
  the latter burst is slightly different from the others,
  cf.\ Fig.~\ref{fig:main_peak}b. It may or may not represent the
  same spectral structure. If it does, the shift of its parameter
  $\tilde{x}$ represents motion of the source relative to the
lens. In  that case we obtain the relation between the lens
  relative velocity  ${v = d\tilde{x}/dt}$ and its size:
  ${a \sim 0.3 \,\mbox{AU} \cdot v / (200 \, 
    \mbox{km}/\mbox{s})}$.

It is worth noting that the narrow bandwidth of the registered FRB spectra
and large cosmological $\nu^{-2}$  dispersion precludes analysis of
another important lens characteristics: the dispersive time delay of
the transmitted FRB signal. The latter depends on frequency in a
nontrivial way, distorting the FRB image in the $t$---$\nu$ plane into
a peculiar recognizable form, cf.~\cite{Clegg:1997ya},
\cite{Cordes:2017eug}, and Fig.~\ref{fig:spectra_2d}.

Overall, the plasma lens hypothesis is very appealing. However, it has
visible inconsistencies. First, the spikes in Fig.~\ref{fig:main_peak}
do not exactly match the main peak slopes and therefore the
theoretical fit. Second, the height of this peak relative to the
nearby spectrum strongly varies from burst to burst,  cf.\ the
bursts~11A and~11D in Fig.~\ref{fig:main_peak}. Third and finally,
some bursts do not have the main peak at all, see the graph 11Q in 
Fig.~\ref{fig:spectrumAC}, as if the lens voluntarily
disappears and then appears again with precisely the same parameters.

Two of the above properties will be explained in the forthcoming
sections. First, the spectra in Fig.~\ref{fig:spectrumAC} include
oscillations, mostly chaotic, at scales below 100~MHz. They 
certainly deform the main peak. Second, we will observe that the
FRB spectra have narrow bandwidth and their central frequency changes
from burst to burst. This makes the 7.1~GHz peak vanish if it is
outside of the signal band.

\section{Wide--band pattern and the progenitor spectrum}
\label{sec:progenitor-spectrum} 
It is remarkable that the FRB spectra of~\cite{Gajjar:2018bth} are
localized in the relatively narrow bands ${\nu_{\mathrm{bw}}
  \sim \mbox{GHz}}$, but their central frequencies differ
  significantly. In fact, the same properties were observed 
  before in the measured spectra of FRB 20121102A
  \citep{2017ApJ...850...76L,  2019ApJ...877L..19G, Gajjar:2018bth,
    Hessels:2018mvq, Majid:2020dwq} and another repeater FRB
  20180916B \citep{Chawla:2020rds, Pearlman:2020sox}. In this
Section we are going to show that the wide--band envelopes of our
  FRB 20121102A spectra are essentially distorted by a spectacular
  propagation phenomenon similar to wide--band
    scintillations. Separating this effect, we will give a strong
  argument that the progenitor spectra themselves are narrow--band 
  and have strongly variable central frequencies.

\begin{figure}
\includegraphics[width=8.6cm]{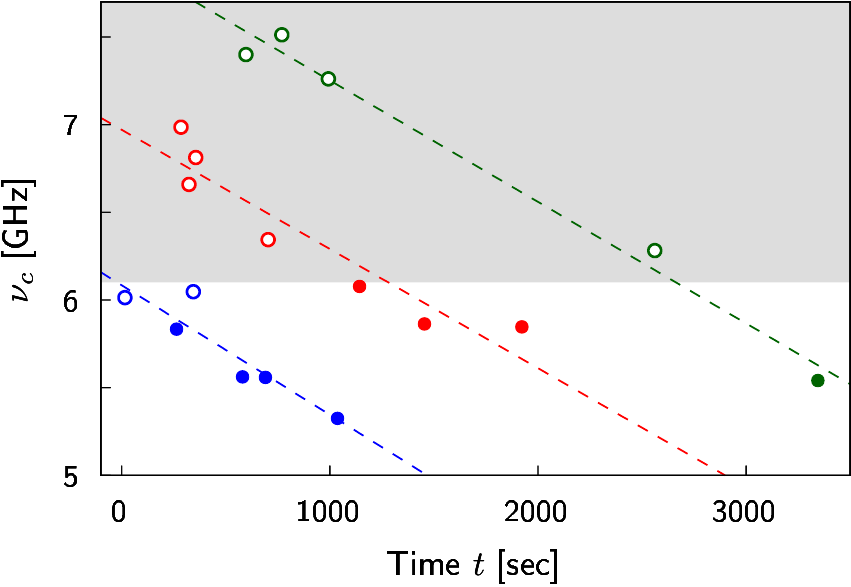}
\caption{
  The central frequencies $\nu_c$ of the bursts versus their arrival
  times $t$. Empty and filled points mark the spectra with the
  pronounced 7.1~GHz peak and without it, respectively. Dashed
  parallel lines are the linear fits of the data from a given
  band. The region $\nu_c >  6.1\; \mbox{GHz}$ is shaded.} 
 \label{fig:LSScintillations}
\end{figure}

To remove the effect of the $7.1\, \mbox{GHz}$ lens, we divide all
registered fluences $F(\nu)$ by the lens gain factor $G(\nu)$
determined from the ``double--peaked'' fit of the spectrum 11D in 
Fig.~\ref{fig:main_peak}b. After that we compute the central 
(``center--of--mass'') frequency of the burst\footnote{Alternatively,
  one may fit the spectra with wide--band parabolas ignoring the data
  between 7.0 and 7.2~GHz. We checked that the positions of the
  parabolic maxima are consistent with $\nu_c$
  in~Eq.~(\ref{eq:CF}).}, 
\begin{equation}
  \nu_c \equiv  \left( \int_{\nu_1}^{\nu_2} \nu \, d\nu \, \frac{F(\nu)}{G(\nu)} \right)
  \Bigg{/}\left( \int_{\nu_1}^{\nu_2} d\nu \, 
  \frac{F(\nu)}{G(\nu)}\right),
  \label{eq:CF}
\end{equation}
where the integrations are performed over the entire signal region
$\nu_1 < \nu < \nu_2$ with positive fluence. It is worth remarking
that Eq.~(\ref{eq:CF}) uses the original unsmoothed fluence
$F(\nu)$.

The central frequency Eq.~(\ref{eq:CF}) of the bursts is plotted in
Fig.~\ref{fig:LSScintillations} as a function of their arrival
time~$t$. Notably, the dependence of $\nu_c(t)$ is not chaotic!
Rather, the central frequencies are attracted to one of
the three parallel inclined bands $\nu_i(t)$ (dashed lines in
Fig.~\ref{fig:LSScintillations}) with seemingly random choice of the band.

The bands at $\nu \approx 6$ and 7~GHz have already been noticed
  by \cite{Gajjar:2018bth} in the summed 4--8 GHz
  spectrum of FRB 20121102. However, their linear evolution with time
  has never been observed. Although both observations are made
    on the basis of limited statistics, they can be tested in the
    future using larger data samples.

Now, Fig.~\ref{fig:LSScintillations} strongly suggests that the
  three bands represent a propagation effect, e.g.\ strong
scintillations of the FRB signal in the turbulent interstellar
plasma. In this case linear time evolution appears due to relative
  motion of the observer and progenitor with respect to the
  scintillating medium. From the physical viewpoint the
scintillations are caused by a refraction of the radio waves on the plasma
fluctuations which makes them propagate via multiple
paths. Interference between the paths 
then distorts the registered FRB spectra into a pattern of
alternating peaks and dips. The bands  
$\nu_i(t)$ in Fig.~\ref{fig:LSScintillations} may represent the scintillation 
maxima. Then the typical distance between them ${\nu_d'  \sim \nu_{i+1} -
\nu_i \sim 0.95\; \mbox{GHz}}$ estimates the decorrelation bandwidth. 

One traditionally characterizes the scintillating plasma with
the diffractive length~--- the typical transverse distance 
  $r_{\mathrm{diff}}$ at which the correlations
  between the radio rays die away. This quantity is related to the 
decorrelation bandwidth as ${\nu_d' / \nu =
    (r_{\mathrm{diff}}'/r_{F,\,S}')^2}$, where $r_{F,\, S}'$ is the 
respective Fresnel scale; see~\cite{Narayan} and
Appendix~\ref{sec:scintillations-1}. We obtain
${r_{\mathrm{diff}}' \sim 0.4 \, r_{F,\, S}'}$. A benchmark  property
of strong diffractive scintillations is an order--one modulation of
the spectra which is observed here, indeed: the regions with strong
signal form isolated islands of GHz width, and fluence between them is
negligibly small, see Fig.~\ref{fig:spectrumAC}.

Importantly, the scintillation pattern is expected to evolve
smoothly with time 
if the source (observer)\footnote{These two options correspond to
  scintillations in the FRB host galaxy and in (some parts of) the
  Milky Way, respectively. We cannot discriminate between them on
    the basis of the spectral data.}
has a nonzero relative velocity $v$ with respect  
to the  scintillating plasma. This is precisely what we see in
Fig.~\ref{fig:LSScintillations}: the maxima $\nu_i(t)$ drift linearly with
the characteristic timescale ${t_{\mathrm{diff}}' \simeq 1350 \,
\mbox{s}}$. Equating ${t_{\mathrm{diff}}' \sim r_{\mathrm{diff}}'/v}$,
we relate the velocity $v$ to ${r_{F,\, S}' = (d'/2\pi \nu)^{1/2}}$ and
hence to the typical distance $d'\sim v_{200}^2 \cdot 2\, \mbox{kpc}$
between the scintillating plasma and the source (observer), where we
introduced ${v_{200} = v/ (200 \, \mbox{km}/\mbox{s})}$.

There remains a question, why the registered spectra have the form of
a single relatively narrow signal region despite the fact that the two or three
scintillation maxima are usually present in the  observation band ${4 \,
  \mbox{GHz} < \nu < 8 \, \mbox{GHz}}$,
cf.\ Figs.~\ref{fig:spectrumAC},~\ref{fig:LSScintillations}.  This can
happen only if the FRB progenitor has a comparably narrow
spectrum with bandwidth $\nu_{\mathrm{bw}} \approx \mbox{GHz}$ which is
capable of ``illuminating'' only one maximum. The same conjecture  explains 
another  feature of Fig.~\ref{fig:LSScintillations}: all bursts with
the main peak (empty circles) have central frequencies within the GHz
band around $7.1\, \mbox{GHz}$ (the shaded region in
Fig.~\ref{fig:LSScintillations}). Note that it would be impossible to
explain the disappearance of the main peak in some bursts by destruction
of the lens: the shape of this peak remains stable prior to
disappearance and recovers later with precisely the same parameters,
cf.~Figs.~\ref{fig:7GHz} and~\ref{fig:LSScintillations}.

To sum up, we have argued that the spectrum of the FRB progenitor
has $\nu_{\mathrm{bw}} \approx \mbox{GHz}$ bandwidth and its
central frequency is changing from burst to burst~--- chaotically or
on short timescale. Note that our argument is based on the separation
of the progenitor properties from the wide--band propagation phenomena
which strongly distort this spectrum with the unknown
frequency shifts of order GHz.

At 1.4 GHz, the registered spectra of FRB 20121102A also occupy a
  narrow band $\nu_{bw}\sim 200$ MHz or ${\nu_{bw}/\nu \sim}$~20\%, as
  was observed by \cite{Hessels:2018mvq}. Note, however, that the
  entire bandwidth of their instrument is comparable to $\nu_{bw}$. In
  that case the  spectral minima at the band boundaries  may be
  provided by the wide--band scintillations
  similar\footnote{But on a different scale, since 
      ${\nu_d' \sim \mbox{GHz}}$ at reference frequency 6~GHz
      corresponds to  
      1.6~MHz at 1.4~GHz according to Kolmogorov law.} to ours. This 
  means that the true bandwidth of the progenitor spectrum may be
  larger at 1.4~GHz: $\nu_{bw} \gtrsim  200\; \mbox{MHz}$. A
    different interesting possibility is that the bandwidth of
  the progenitor spectrum always constitutes 20\% of its central frequency. But
  this latter assumption still has to be tested with the wide--band
  measurements.

Despite distortions, one can search for the periodic evolution of the
progenitor central frequency $\nu_c(t)$. We performed this search
using the periodogram method described in~\cite{Zechmeister:2009js}
and~\cite{Ivanov:2018byi} at time scales $60 - 1000 \,
  \mbox{s}$. The best--fit value for a period is 112~s,   but the
effect is not statistically significant.

\section{Narrow--band scintillations}
\label{sec:scintillations}
At shorter scales $\Delta \nu \lesssim 100 \, \mbox{MHz}$ the
spectra in Fig.~\ref{fig:spectrumAC} display seemingly chaotic pattern
of alternating peaks and dips. It would be natural to explain this
random behavior with another, narrow--band kind of strong interstellar
scintillations. Let us show that the latter are indeed present in the
FRB 20121102A spectra.

\begin{figure}
\includegraphics[width=8.6cm]{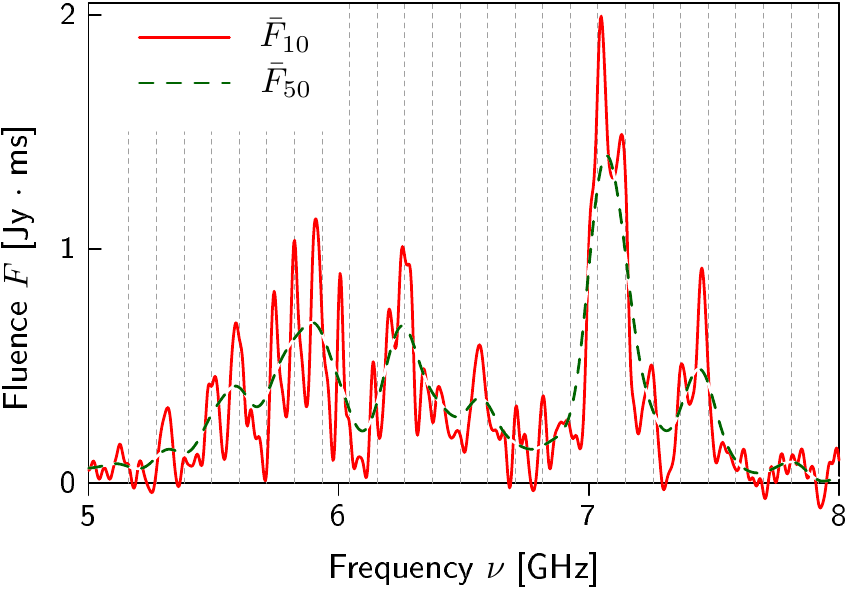}
\caption{Fluence $F(\nu)$ of the burst 11A smoothed with windows
  ${\sigma=10\;   \mbox{MHz}}$ (solid line) and $50\;\mbox{MHz}$
  (dashed). Thin vertical lines display uniform lattice with spacing
  ${T_\nu = \mbox{110.3 MHz}}$.}
 \label{fig:periodic_11A}
\end{figure}
It is natural to treat the scintillations statistically, i.e.\ average the
spectra over a large ensemble of turbulent plasma clouds and then
compare the mean observables to
the theory. We recall~\citep{Rickett1990, Narayan, Lorimer-Kramer,
  Woan} that different-frequency waves refract differently
and therefore go along 
different paths though statistically independent volumes of the
turbulent medium. This makes the scintillating radio spectra
uncorrelated at frequencies $\nu$ and $\nu+\Delta \nu$ if $\Delta
\nu$ exceeds the decorrelation bandwidth $\nu_d$. As a consequence,
the statistical average can be performed by integrating over many $\nu_d$
intervals. Below we regard the fluence $\bar{F}_{50}(\nu)$ smoothed
with large window $\sigma = 50 \; \mbox{MHz}$ in Eq.~(\ref{eq:2}) as a
statistical mean. This quantity indeed delineates a wide--band
envelope of the spectrum in Fig.~\ref{fig:periodic_11A} (dashed line)
with no trace of the erratic short--scale structure. Note, however, that
$\bar{F}_{50}$ should be interpreted with care, since smoothing in
Eq.~(\ref{eq:2}) destroys any oscillatory behavior, erratic or not, at
frequency periods below $\pi \sigma \sqrt{2} \sim 220\; \mbox{MHz}$.

We introduce the autocorrelation function characterizing the
statistical dependence of the spectral fluctuations ${\delta F =
  F -  \bar{F}_{50}}$ at frequencies $\nu$ and $\nu + \Delta \nu$, 
\begin{equation}
  \label{eq:25}
  \mbox{ACF}(\Delta \nu) = {\cal N}\int\limits_{\nu_1}^{\nu_2-\Delta \nu} d\nu \;\;
  \frac{\delta F(\nu) \delta F(\nu+\Delta \nu)}{\nu_2 - \nu_1 - \Delta \nu}\;,
\end{equation}
where the data are averaged over the signal bandwidth $\nu_1 < \nu
  < \nu_2$ by integrating and dividing by the integration
  interval, whereas the normalization factor ${\cal N}$ makes
$\mbox{ACF}(0)= 1$. We will see that Eq.~(\ref{eq:25}) is a convenient 
observable sensitive both to scintillations and periodic structures in
the spectra. 

\begin{figure}
  \unitlength=1mm
  \begin{picture}(86,80)
    \put(0,0){\includegraphics[width=8.6cm]{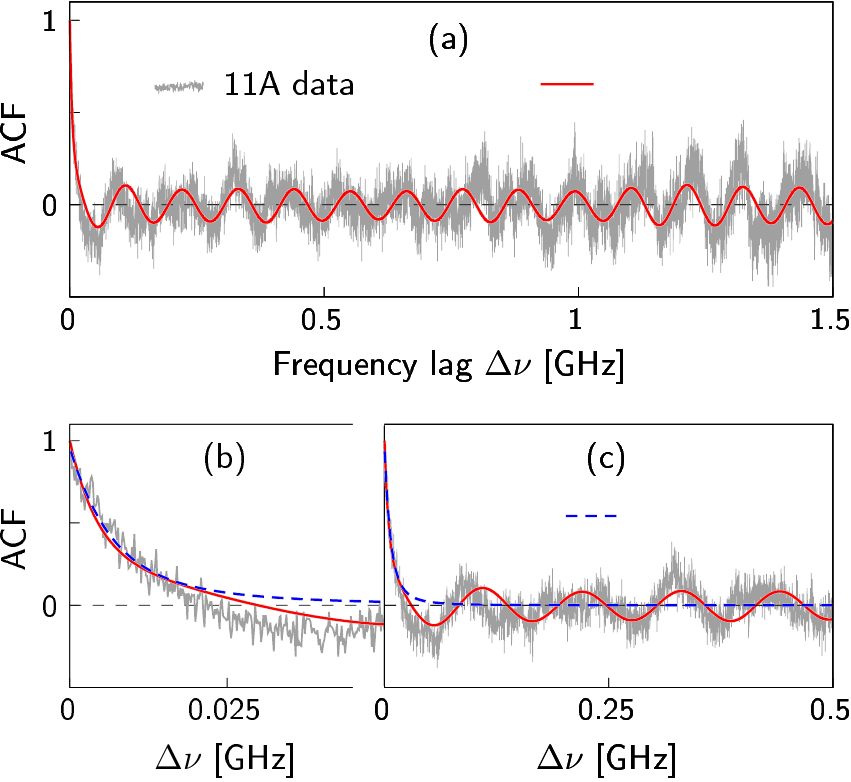}}
    \put(61.8,69.3){\fontsize{10}{1}\sffamily Eq.~(\ref{eq:59})}
    \put(64.1,25.6){\fontsize{10}{1}\sffamily Eq.~(\ref{eq:55})}
  \end{picture}
\caption{(a) The autocorrelation function Eq.~(\ref{eq:25}) of the burst 11A.
  (b), (c) Its zoom--ins at small $\Delta \nu$. Dashed and solid
  lines show fits by the theoretical models describing
  scintillations and scintillations+lensing, Eqs.~(\ref{eq:55}) and
  (\ref{eq:59}), respectively.}
\label{fig:ACF}
\end{figure}

In Fig.~\ref{fig:ACF} we plot $\mbox{ACF}(\Delta \nu)$ for the burst
11A. It decreases at first indicating that $\delta F(\nu)$ and $\delta
F(\nu+\Delta \nu)$ are less correlated at larger $\Delta \nu$. But
surprisingly, at $\Delta \nu \gtrsim 50\; \mbox{MHz}$ the
  autocorrelation function develops a set of wide almost equidistant
maxima suggesting that the coherence partially returns! We will
consider this effect in the next Section. 

\begin{figure}
  \unitlength=1mm
  \centerline{
    \begin{picture}(60,40)
      \put(0,0){\includegraphics{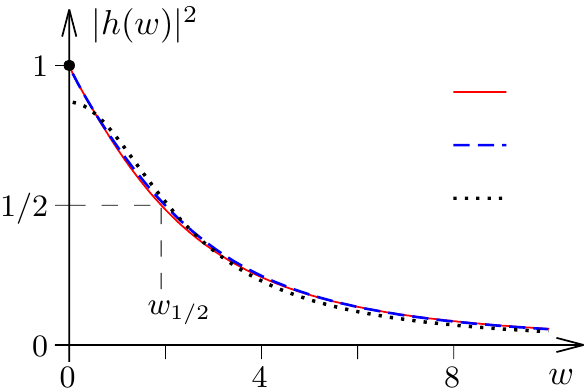}}
      \put(32,29.3){\normalsize \sffamily theory}
      \put(32,24){\normalsize \sffamily Eq.~(\ref{eq:75})}
      \put(32,18.7){\normalsize \sffamily Eq.~(\ref{eq:36})}
  \end{picture}}
  \caption{Theoretical autocorrelation function $|h(w)|^2$ describing
    Kolmogorov--type scintillations, $w = 2\Delta
    \nu/\nu_d$; see Eq.~(\ref{eq:53}) of
      Appendix~\ref{sec:scintillations-1}. It reaches $1/2$ at
    ${w_{1/2} \approx 1.9}$. Dashed and dotted lines show the
      approximation Eq.~(\ref{eq:75}) and the Lorentzian
      fit Eq.~(\ref{eq:36}), respectively.}
  \label{fig:h}
\end{figure}
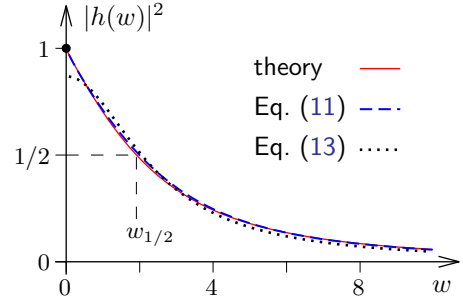
To interpret the data, we theoretically computed the autocorrelation
function for the radio waves strongly refracted in the turbulent plasma 
with the standard Kolmogorov--type distribution of free electrons, see
Appendix~\ref{sec:scintillations-1}. The result\footnote{It is worth noting
that the analytic formula~(\ref{eq:55}) is valid at ${\nu_d \ll
\nu}$ with corrections of order ${(\nu_d/\nu)^{1/3} \sim 8\%}$.} is,
\begin{equation}
  \label{eq:55}
  \mbox{ACF} = {\cal N} \int\limits_{\nu_1}^{\nu_2 - \Delta \nu} 
  \frac{\bar{F}_{50}^2(\nu)\; d\nu}{ \nu_2 -
    \nu_1 - \Delta \nu} \;
\left| h\left(\frac{2\Delta \nu}{\nu_d(\nu)}\right)\right|^2\;, 
\end{equation}
where $\bar{F}_{50} \approx \langle F \rangle$  again substitutes the
statistical average, while $|h(w)|^2$ is a universal
hat--like function depicted with the solid line in Fig.~\ref{fig:h}.
Strictly speaking, $h$ is given by the
integral~(\ref{eq:53}), but in 
practice one can use a very good approximation
   \begin{equation}
     \label{eq:75}
     h(w) \approx [1+ a (iw)^{5/6} + (iw/b)^{7/4}]^{-4/7}
   \end{equation}
capturing the small-- and large--$w$ asymptotics of this function and
  therefore correctly representing it at the intermediate values as well, see
the dashed line in
Fig.~\ref{fig:h}. In Eq.~(\ref{eq:75}) we used the
   numerical coefficients ${a = \frac78 \Gamma(11/6)}$,
   $b=\Gamma(11/5)\, 2^{6/5}$ and Euler gamma--function 
   $\Gamma(z)$.

The only fitting parameter of the theoretical model Eq.~(\ref{eq:55}) is the value
of the decorrelation bandwidth $\nu_d(\nu)$ at a given frequency, say,
$\nu_{d,\, 6} \equiv \nu_d(6\; \mbox{GHz})$. At other frequencies the
bandwidth is determined from the Kolmogorov scaling
\begin{equation}
  \label{eq:54}
    \nu_d(\nu) =  \nu_{d,\, 6}
    \left(\frac{\nu}{6\, \mbox{GHz}}\right)^{4.4} = \nu \;
  (r_{\mathrm{diff}}/r_{F,\, S})^2\;.
\end{equation}
The first of these equations is convenient in practice, while the
second relates $\nu_d$ to the parameters of the scintillating medium:
the Fresnel scale  $r_{F,\, S}$ characterizing its  distancing from the
source or the observer and diffractive length $r_{\mathrm{diff}}$~---
the transverse separation at which the radio paths decohere inside the
medium, see Eqs.~\eqref{eq:56} and~(\ref{eq:9}).

We stress that the theoretical expression~(\ref{eq:55}),
  (\ref{eq:75}) is new. Previous studies of~\cite{Hessels:2018mvq}
  and~\cite{Majid:2020dwq} traditionally assumed a Lorentzian ACF
  profile\footnote{\cite{Gajjar:2018bth} used a Gaussian profile
      which does not resemble our theoretical ACF.},
  \begin{equation}
    \label{eq:36}
    h_L^2 = {\cal N} \left[c + (\Delta \nu^2 + \nu_{d,\, L}^2)^{-1}\right]\;,
  \end{equation}
  which has two parameters: the bandwidth $\nu_{d,\, L}$ and an
  additive constant $c$. At $c=0$, the fit of our theoretical prediction
  $|h(w)|^2$ with this function gives the dotted line in
  Fig.~\ref{fig:h} and ${\nu_{d,\, L} \approx 1.2\,\nu_d}$. Notably, the
  Lorentzian profile works very well at intermediate frequency lags
  but, being regular at $\Delta \nu\to 0$, deviates from the theory at
  small~$w$. As a consequence,  even at $c=0$ it underestimates the
  height of the ACF, overestimates its half--height width and gives
  20\% larger value of~$\nu_d$. We will see below that the
  two--parametric Lorentzian fits with arbitrary $c$ are much worse.

Importantly, our theoretical $|h|^2$ decreases with $\Delta \nu$
from ${h(0)=1}$ to zero reaching $1/2$ at $\Delta \nu \approx 0.96\,
\nu_d$.  The full autocorrelation functions Eq.~(\ref{eq:55}) have 
similar profiles and, in particular, monotonically fall off at
large $\Delta \nu$. This is the 
only possible behavior because random fluctuations in the turbulent plasma
suppress correlations between the different--frequency waves, and the
suppression is stronger at larger~$\Delta \nu$. As a consequence,
Eq.~(\ref{eq:55}) (dashed line) fits  well the initial falloff of the
experimental ACF in Fig.~\ref{fig:periodic_11A}b giving $\nu_{d,\, 6}
= 3.5\; \mbox{MHz}$. But the same theory fails to
explain the peaks  at larger~$\Delta \nu$ which will be considered
in the next Section.

 \begin{figure}
   \centerline{\includegraphics{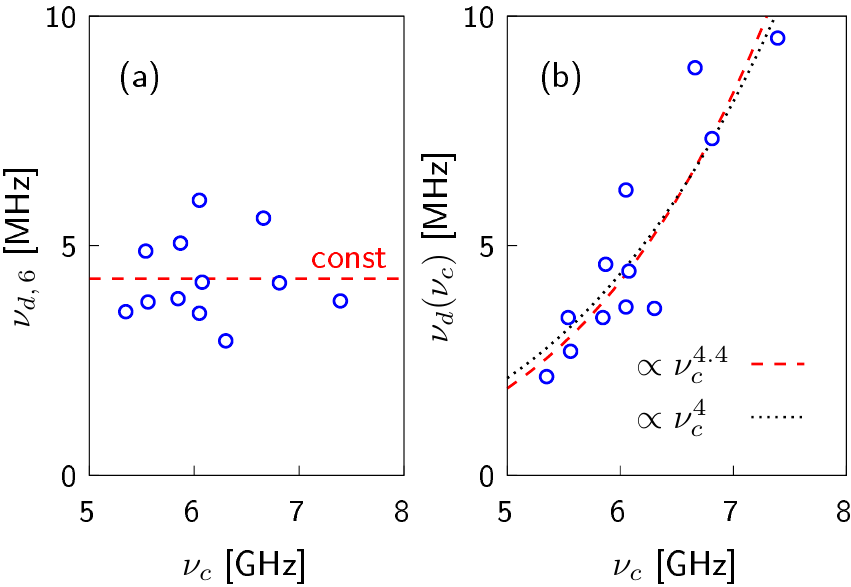}}
   \caption{(a) Decorrelation bandwidths $\nu_{d,\, 6} = \nu_d(6\;
     \mbox{GHz})$ of the burst spectra versus their central
     frequencies $\nu_c$. (b)~Bandwidths $\nu (\nu_c)$ rescaled to
     $\nu = \nu_c$ via Eq.~(\ref{eq:54}). Dashed lines
     show fits by  Eq.~(\ref{eq:54}): ${\nu_d \propto \nu_c^{4.4}}$
     and $\nu_{d,\,  6} = \mbox{const}$. Dotted line in the right
     panel demonstrates weaker frequency dependence $\nu_d \propto
     \nu_c^{4}$.}
   \label{fig:nud}
 \end{figure}
Overall, we find that the autocorrelation functions of the 12 most powerful
bursts match Eq.~(\ref{eq:55}) at small $\Delta \nu$ whereas the
weaker bursts 11B, C, G, J, K, M are dominated by the instrumental noise
and do not produce discernible correlation patterns at all. From these 
fits we obtain 12 values of $\nu_{d,\, 6}$ in Fig.~\ref{fig:nud}a. The
data points group around a constant
\begin{equation}
  \label{eq:15}
  \nu_d(6~\mbox{GHz})  =  4.3\pm 0.9\; 
  \mbox{MHz}
\end{equation}
despite the fact that their spectra  
are localized in essentially different frequency regions. Rescaling the spectral
bandwidths $\nu_d$ to the burst central frequencies $\nu = \nu_c$ 
via Eq.~(\ref{eq:54}), we obtain the function $\nu_d(\nu_c)$ in
Fig.~\ref{fig:nud}b which closely follows the Kolmogorov scaling
(dashed line).

The mean value of $\nu_d$ fixes the parameter
${r_{\mathrm{diff}}/r_{F,\, S} \approx \mbox{0.027} \pm 0.003}$
of the scintillating plasma. Assuming a galactic distance $d$ to it,
we obtain a reasonable diffractive length ${r_{\mathrm{diff}}\sim (1.3
  \pm 0.1) \times 10^{9}\, \mbox{cm} \times  (d/\mbox{kpc})^{1/2}}$,
cf.~Eq.~(\ref{eq:9}). 

It is worth noting that the frequency integral is an important part of
Eq.~(\ref{eq:55}) because the decoherence bandwidth $\nu_d$ strongly
depends on $\nu$. The theoretical result simplifies, however, in
  the case $\nu_2 - \nu_1 \ll \nu$ when
\begin{equation}
  \label{eq:62}
  \mbox{ACF} = \left| h(2\Delta \nu/ \nu_d)\right|^2
\end{equation}
is completely determined by the hat--like function in Eq.~(\ref{eq:75}),
while $\nu_d$ is computed either at $(\nu_2 + \nu_1)/2$ or at $\nu_c$
if the spectrum itself is narrow--band. The price to pay, however, is larger
statistical fluctuations in the smaller data set. Since our data
  are relatively narrow--band, we fitted the ACF's of the 12 strongest
  bursts by Eq.~(\ref{eq:62}). The respective values of $\nu_d(\nu_c)$
  were consistent with Fig.~\ref{fig:nud}b. Using them in
  Eq.~(\ref{eq:54}), we  arrived to ${\nu_{d}(6\;\mbox{GHz})  = 4.0 \pm
  0.8\, \mbox{GHz}}$~--- almost the same
  result as before.

\begin{figure}
  \centerline{\includegraphics[width=8.6cm]{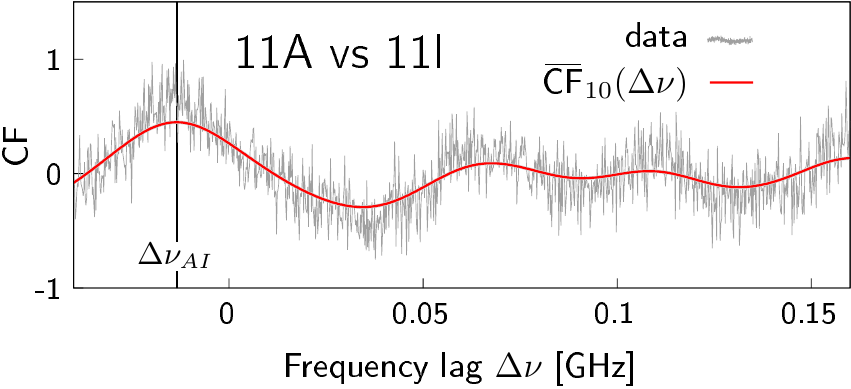}}
  \caption{Correlation function $\mbox{CF}(\Delta \nu)$ between the
      bursts 11A and 11I. To determine its highest maximum (vertical
      line), we Gauss--smooth this function with window
      ${\sigma = 10\; \mbox{MHz}}$ (smooth solid line).}
  \label{fig:cf}
\end{figure}
\begin{figure}
\includegraphics[width=8.6cm]{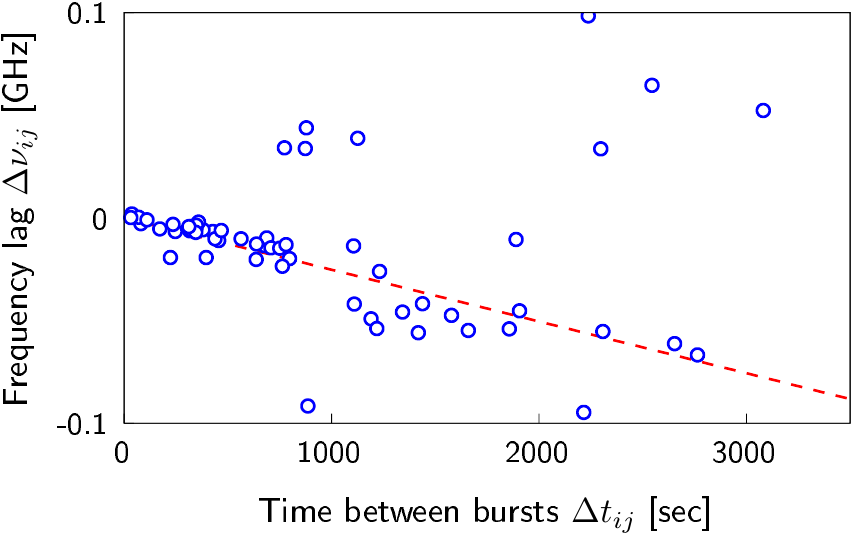}
\caption{Frequency shifts $\Delta \nu_{ij}$ between the pairs of the
  burst spectra versus the time $\Delta t_{ij}$ elapsed between them.}
 \label{fig:SSScintillations}
\end{figure}
We finish this Section with an extra argument, why erratic
narrow--band structure stems from a propagation
effect, and it is not just a stochastic variation of the
  progenitor spectrum. First, we  
compute the correlation functions (CFs) between the pairs of the
burst spectra by substituting their fluences ${\delta F_i(\nu) \equiv F_i - 
\bar{F}_{50,\, i}}$ and $\delta F_j (\nu)$ into Eq.~(\ref{eq:25})
instead of the two identical $\delta F$'s. In particular, in Fig.~\ref{fig:cf} we
plot $\mbox{CF}(\Delta 
\nu)$ between the bursts 11A and 11I. The highest peak of 
this function occurs at nonzero $\Delta \nu = \Delta \nu_{AI}$
suggesting that the erratic structures in the spectra are 
shifted with respect to each other. Moreover, the frequency shift
$\Delta \nu_{ij}$ between the different burst pairs linearly depends
on the
time elapsed between them, see\footnote{To increase
  the significance, in Fig.~\ref{fig:SSScintillations} we consider only
  the bursts with relatively close central frequencies, ${|\Delta
    \nu_{c}| <  0.5 \; \mbox{GHz}}$.} Fig.~\ref{fig:SSScintillations}:
${\partial_t \Delta\nu_{ij}\approx -2.52\cdot 10^{-2} \;
  \mbox{MHz}/\mbox{s}}$ (dashed line). Like in the previous Section,
we explain this effect by a
relative motion of the observer (source) with respect to the
scintillating medium~\citep{Rickett1990, Narayan, Lorimer-Kramer,
  Woan}. The velocity is then roughly estimated as $ v\sim
r_{\mathrm{diff}} \, \partial_t\Delta \nu_{ij} /\nu_d \sim 76 \, 
  \mbox{km}/\mbox{s} \; (d/\mbox{kpc})^{1/2}$, where the experimental
values for $r_{\mathrm{diff}}$ and $\nu_d$ are substituted. We obtained 
a reasonable galactic velocity.

\section{Periodic structure}
\label{sec:periodic-structure}
\begin{figure}
  \centerline{\includegraphics{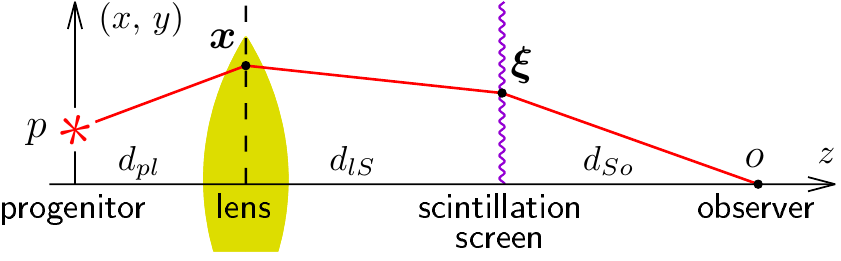}}
  \caption{A radio ray passing through the lens and the scintillating medium.}
  \label{fig:scintillation_screen}
\end{figure}

So far we have argued that the peaks of the autocorrelation
function in Fig.~\ref{fig:ACF} cannot originate from the scintillations
because the latter introduce stronger 
suppression at larger $\Delta \nu$. The same peaks, however, are
naturally  explained by the wave diffraction. Indeed, suppose that before
or after hitting the scintillating medium the FRB signal passes
through the lens~--- a plasma cloud or a vicinity of a compact
gravitating body~--- which splits it into two radio rays,
\begin{equation}
  \label{eq:63}
  f = f_p \left( G_{1}^{1/2} \,\mathrm{e}^{i\Phi_1}  +  G_{2}^{1/2}
  \,\mathrm{e}^{i\Phi_2} \right) \;,
\end{equation}
see one of these rays in Fig.~\ref{fig:scintillation_screen}. We
introduced the intrinsic progenitor signal $f_p(\nu)$, gain factors 
$G_{1,\,2}$ of the rays, and  their phases $\Phi_{1,\, 2}$.  As a
consequence of  Eq.~(\ref{eq:63}), the net  FRB fluence $F = |f|^2$
includes an interference term proportional to $\cos(\Phi_1 - \Phi_2)$
which oscillates as a function of frequency with the
period ${T_\nu = 2\pi |\partial_\nu(\Phi_2 -  \Phi_1)|^{-1}}$. This
enhances correlations between $F(\nu)$ and ${F(\nu + \Delta \nu)}$
at frequency lags equal to the multiples of the frequency period,
${\Delta \nu = nT_\nu}$ and therefore produces equidistant peaks in
the autocorrelation function Eq.~(\ref{eq:25}).

Scintillations obscure the above picture by adding a random component
to the wave Eq.~(\ref{eq:63}). In Appendix~\ref{sec:scintillations-1} we
develop an analytic model for the radio waves propagating through the
lens and the scintillating plasma. The spectral fluence in this case
equals (cf.\ Eq.~(\ref{eq:40})),
\begin{multline}
  \label{eq:66}
     F  = \bar{F}_{50}\left[1 + A_{\mathrm{osc}} \; 
       \mathrm{e}^{ - \frac12
         (|\tilde{\bm{\xi}}_1-\tilde{\bm{\xi}}_2|/r_{\mathrm{diff}})^{5/3}}
         \cos(\Phi_1 - \Phi_2)\right]\\ + \delta' F \;,
\end{multline}
where we extracted a smooth envelope $\bar{F}_{50}(\nu)$ of the
spectrum, denoted the relative oscillation amplitude by
${A_{\mathrm{osc}} \equiv 2\sqrt{G_1G_2}/(G_1 + G_2)^{-1} < 1}$, and
introduced the transverse distance ${|\tilde{\bm{\xi}}_{2} -
  \tilde{\bm{\xi}}_1|}$ between the radio rays inside the
scintillating medium. The first line in Eq.~(\ref{eq:66})
represents the statistically averaged contributions of the radio rays
and their interference, whereas $\delta' F$ denotes fluctuations
  of the fluence. Scintillations suppress the interference term if 
the rays go too far apart. We are interested in the unsuppressed
regime\footnote{At $T_\nu \ll   \nu_d$ the interference can be
  registered even if  
  Eq.~(\ref{eq:67}) is broken, see the discussion in
  Appendix~\ref{sec:two-radio-paths}. However, Fig.~\ref{fig:ACF}
  suggests $T_{\nu} > \nu_d$, so we disregard this possibility.} 
\begin{equation}
  \label{eq:67}
  |\tilde{\bm{\xi}}_{2} - \tilde{\bm{\xi}}_1|
    < r_{\mathrm{diff}}\;.
\end{equation}
Notably, in Appendix~\ref{sec:scintillations-1} we find out that this
inequality easily holds for scintillations occurring in our galaxy if the
lens is outside of it, cf.\ Eq.~(\ref{eq:58}). Or vice versa: one may
imagine that the scintillations happen in the FRB
host galaxy and the lens is either in the intergalactic space or in
the Milky Way.

But even if Eq.~(\ref{eq:67}) holds, the interference peaks of
$F(\nu)$ are hidden in the sea of erratic fluctuations $\delta'
F$. The latter have order--one amplitude if the scintillations are strong:
$\delta' F \sim F$, cf.\ Fig.~\ref{fig:periodic_11A}. To separate
the two effects, one uses the autocorrelation
function Eq.~(\ref{eq:25}) where
the frequency integral averages the fluctuations
away. In Appendix~\ref{sec:scintillations-1} we derive ACF for the
theoretical model with scintillations$+$lensing, 
\begin{multline}
  \label{eq:59}
  \mbox{ACF} = {\cal N} \int_{\nu_1}^{\nu_2 - \Delta \nu} d\nu \;
  \frac{\bar{F}_{50}(\nu) \bar{F}_{50}(\nu + \Delta \nu)}{ \nu_2 -
    \nu_1 - \Delta \nu} \; \\ \times \Big[ \,|h|^2  +\frac12 A_{\mathrm{osc}}^2\, \left(
  1 + |h|^2\right) \, \cos(2\pi \Delta \nu/T_\nu) \Big]\;,
\end{multline}
where $h\equiv h(2\Delta \nu/\nu_d(\nu))$ is the same ``scintillation''
function~(\ref{eq:75}) as before, $\nu_d(\nu)$ is given by
Eq.~(\ref{eq:54}), and the frequency period
$T_{\nu}$ is already extracted from $\Phi_2 - \Phi_1$. Note that
Eq.~(\ref{eq:59}) is valid at $\nu_d \ll \nu$ with the corrections of
order $(\nu_d/\nu)^{1/3} \sim 8\%$.   

The theoretical expression~(\ref{eq:59}) describes both the
initial falloff of the autocorrelation function due to
scintillations and the periodic peaks  at ${\Delta \nu = nT_{\nu}}$
caused by the two--ray interference. It fits well the
experimental data in Fig.~\ref{fig:ACF} (solid
line) giving $\nu_d \approx 2.4\; \mbox{MHz}$, ${T_{\nu} \approx
  110.3\; \mbox{MHz}}$, and $A_{\mathrm{osc}} \approx 0.5$ for
  the burst 11A.

\begin{figure}
  \unitlength=1mm
  \begin{picture}(87,40)
    \put(0,0){\includegraphics[width=8.6cm]{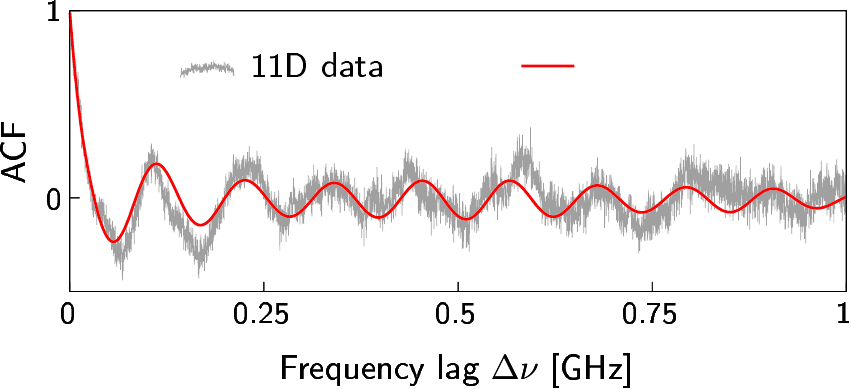}}
    \put(60,31.7){\sffamily Eq.~(\ref{eq:59})}
  \end{picture}
  \caption{Autocorrelation function of the burst 11D.}
  \label{fig:acf_11D}
\end{figure}

 \begin{figure}
   \centerline{\includegraphics{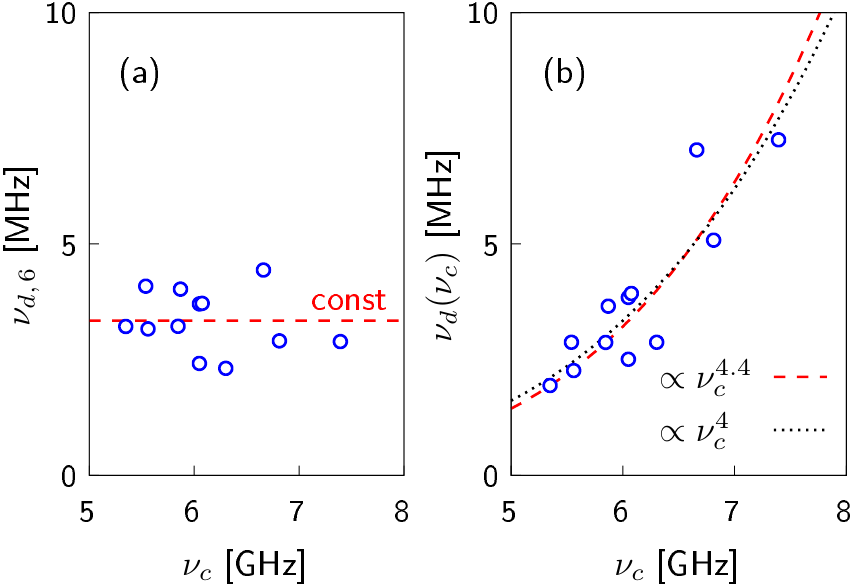}} 
   \caption{(a)~Decorrelation bandwidths ${\nu_{d,\, 6} \equiv
     \nu_d(6\; \mbox{GHz})}$ of the burst spectra extracted from
       the full fits Eq.~(\ref{eq:59}) of their ACFs that take into
     account the periodic structure. (b)~Bandwidths $\nu_{d}(\nu_c)$
     rescaled to the central frequencies of their bursts via
     Eq.~(\ref{eq:54}). Dashed lines represent the mean bandwidth
       ${\nu_{d,\,  6} \approx 3.3\pm 0.6
       \,\mbox{MHz}}$ rescaled with Eq.~(\ref{eq:54}). The dotted line
     shows weaker dependence $\nu_d \propto \nu_c^{4}$ for
     comparison.}
   \label{fig:nud_periodic}
 \end{figure}

 \begin{figure}
  \centerline{\includegraphics[width=8.6cm]{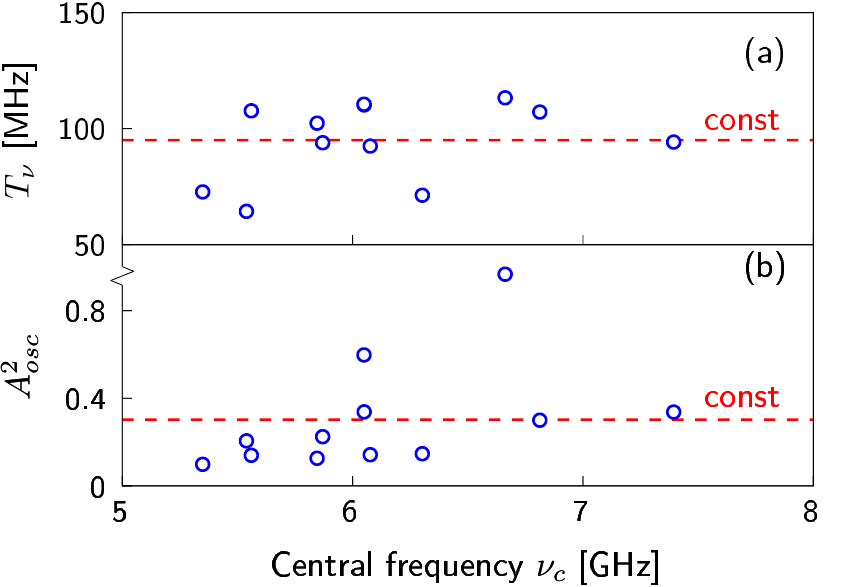}}
  \caption{(a) Frequency periods $T_\nu$ and (b) amplitude squares
    $A_{\mathrm{osc}}^2$ of the spectral oscillations in bursts with
    central frequencies~$\nu_c$.}
  \label{fig:Tnu}
\end{figure}

In fact, the autocorrelation functions of the strongest bursts, e.g.\ 11A
or 11D, display easily recognizable sets of periodic peaks, see
Fig.~\ref{fig:acf_11D}, and many other bursts include hints of 
those. Fitting ACFs of the 12 most powerful
spectra\footnote{\label{powerful}The same as before i.e.\ all
  except 11B, C, G, J, K,~M.} with Eq.~(\ref{eq:59}), we obtain
the respective decorrelation bandwidths $\nu_d$, frequency
periods $T_{\nu}$, and amplitudes $A_{\mathrm{osc}}$ in\footnote{The
  periods $T_{\nu}$ and central frequencies $\nu_c$ of the bursts 11A
  and 11E are indistinguishably close to each other in
  Fig.~\ref{fig:Tnu}a. We will comment on this feature below.} 
Figs.~\ref{fig:nud_periodic} and~\ref{fig:Tnu}.

The frequency periods in Fig.~\ref{fig:Tnu}a group within the $15\%$ interval
around the mean value $T_\nu \approx 95\pm 16\; \mbox{MHz}$ and do not
indicate any dependence on frequency. At the
same time, the jumps of $A_{osc}^2$ in 
Fig.~\ref{fig:Tnu}b are much larger. This sensitivity can be explained
by the fact that some weak ACF's  include only hints of the periodic
patterns and give very small $A_{osc}$, while the other have barely
discernible initial ``scintillation'' falloffs, hence
an overestimate of~$A_{osc}$. In what follows we use $A_{osc}
\approx 0.5$ obtained by averaging the 12 points in
Fig.~\ref{fig:Tnu}b (dashed line).

\begin{figure}
  \centerline{\includegraphics[width=8.6cm]{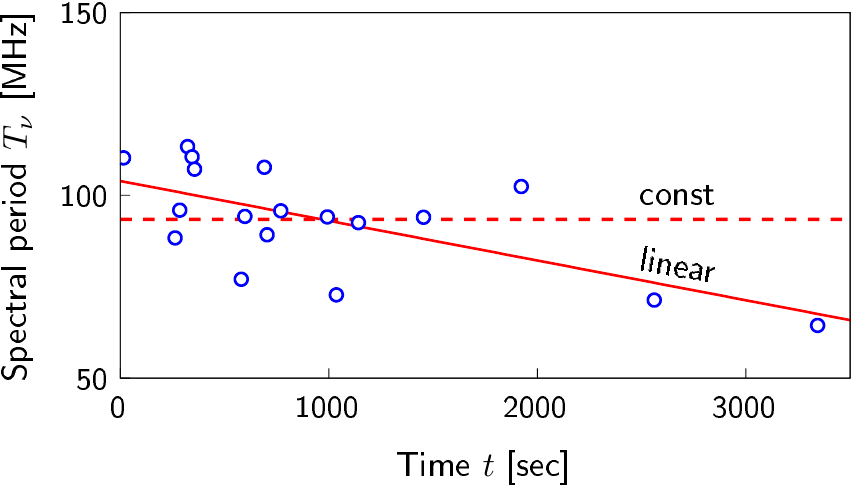}}
  \caption{Periods $T_{\nu}$ of spectral oscillations in different
      bursts versus their arrival times (points). Solid and dashed
      lines show fits by linear time evolution and a constant, 
    respectively.}
  \label{fig:period}
\end{figure}
In Fig.~\ref{fig:period} we plot the frequency periods of different
bursts versus their arrival times. The data points are consistent with the
time--independent $T_{\nu}$ (dashed line), although a slow evolution
with ${dT_\nu/dt \approx - 1.1\cdot 10^{-2} \, \mbox{MHz}/\mbox{s}}$ 
is also possible (solid line). 

The fit Eq.~(\ref{eq:59}) gives an improved estimate of the burst
scintillation bandwidths $\nu_{d,\, 6}$ and $\nu_d(\nu_c)$, see
Figs.~\ref{fig:nud_periodic}a,b and recall Fig.~\ref{fig:nud}. Now,
the data group a bit closer, and they still agree with  the
  prediction of the Kolmogorov turbulence Eq.~(\ref{eq:54}) (dashed
lines). The averaging gives ${\nu_{d}(6\; \mbox{GHz}) \approx
    (3.3\pm 0.6)\, \mbox{MHz}}$. 

Now, we can explicitly visualize the periodic pattern in the burst
spectra. In Fig.~\ref{fig:periodic_11A} we plotted  the 
smoothed fluence $\bar{F}_{10}(\nu)$ of the burst 11A together with
the lattice of the vertical dotted lines separated by ${T_{\nu}=110.3
\; \mbox{MHz}}$~--- a frequency period of the respective ACF in
Fig.~\ref{fig:ACF}. By construction,  smoothing with ${\sigma=10\,
\mbox{MHz}}$ kills almost all narrow--band scintillations because 
${\nu_d \ll \pi\sigma \sqrt{2}}$. As a consequence, the highest
maxima in Fig.~\ref{fig:periodic_11A} should belong to the periodic
structure. Indeed,  too many of them are close to the dotted
lines. Moreover, the side spikes of the ``main'' 7.1~GHz peak seem to
be a part of the same structure.

We demonstrated that the two--ray interference correctly describes
  the leading periodic behavior of the spectra. Nevertheless, there
  are visible inconsistencies. First, the experimental
  autocorrelation functions in Figs.~\ref{fig:ACF}a, \ref{fig:acf_11D} 
  deviate from the fits in some places. Second, there are
  burst--to--burst variations in ACFs leading to 15\% spread of 
  $T_\nu$ values in Fig.~\ref{fig:Tnu}a. Third,  some maxima in 
  Fig.~\ref{fig:periodic_11A} are off the periodic grid.

All these
  effects are expected. On the one hand, strong erratic GHz-- and
  MHz--scale scintillations exist in the spectra; without any  doubt,
  the ones with $\nu_d  \sim 100$~MHz are present as well. They
  slightly shift the maxima in Fig.~\ref{fig:periodic_11A} and add
  smaller  peaks. Moreover, unlike  the strongest narrow--band
  fluctuations which are averaged via the frequency  integral in
  Eq.~(\ref{eq:25}), the ones with larger $\nu_d $ remain almost
  random. They stochastically distort the expected cos--like behavior
  of  the ACFs in Figs.~\ref{fig:ACF}a, \ref{fig:acf_11D}  and
  penetrate into  the fit results for $T_\nu$. On the other hand, we use
  the simplest two--wave interference model and disregard the
  subdominant rays altogether. But generically, the latter are present
  in the data along with their subdominant interference contributions
  distorting the graphs. One can take these features into account at
  the cost of adding new parameters to the fits.  

In fact, the structure similar to what we see in
  Fig.~\ref{fig:periodic_11A} was observed in the High--Frequency
  Interpulses (HFI) of the Crab pulsar, see
  \cite{Hankins:2016bbl}. Namely, the spectra of HFI consist of 
  many isolated bands with inter--band distance ${\nu_{\mathrm{band}}
  \simeq 0.06\, \nu}$. In turn, every band includes $\sim 3$ 
  sub--bands. At 6 GHz this gives ${\nu_{\mathrm{band}} \simeq 344}$ MHz
  and $\sim 100$~MHz of sub-band distance.  Intriguingly, the 
  wide--band envelope $\bar{F}_{50}$ of the FRB spectrum in
  Fig.~\ref{fig:periodic_11A} has several maxima separated by 
  $\nu_{\mathrm{band}} \simeq 330$ MHz which consist, at a higher
  spectral resolution, of the equidistant peaks with period $T_\nu
  \approx 110$~MHz. This resemblance may suggest similar 
  mechanisms operating in Crab and in FRB 20121102A. Note that
  $T_\nu \propto \nu$ does not contradict to the points in
  Fig.~\ref{fig:Tnu}a which have a spread. 

\begin{figure}
\includegraphics[width=8.6cm]{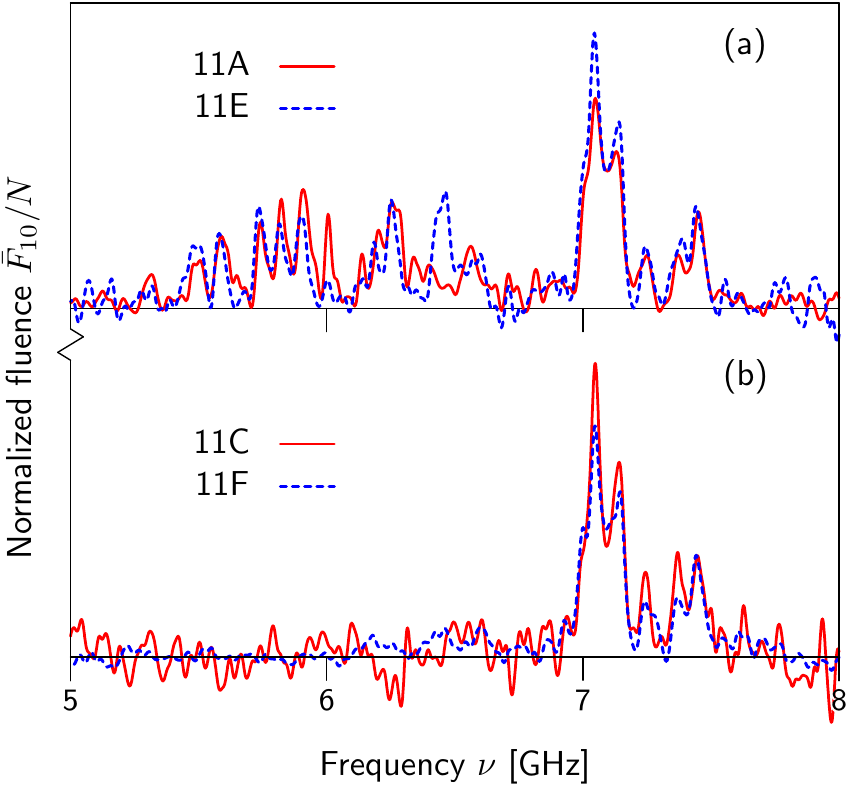}
\caption{Normalized fluences of the (a) bursts 11A and 11E; (b)
    bursts 11C and 11F. The area under each graph equals one.}
\label{fig:spectrumAE}
\end{figure}
Let us argue that the entire 100\; MHz spectral pattern including the
periodic structure and subleading peaks originates from the propagation 
phenomena. We divide the smoothed spectra $\bar{F}_{10}(\nu)$
by their total energy releases making the areas under their graphs
equal to one. After that  some of the normalized spectra look almost
identical to each other, like the twin brothers, cf.\ the graphs 11A
and 11E in Fig.~\ref{fig:spectrumAE}a, or 11C and 11F in
Fig.~\ref{fig:spectrumAE}b. Notably, these coinciding bursts are not
sequential, e.g.\ 11A is followed by the bursts B to D, and only then
by the burst E. Such similarity would be very hard to explain by the intrinsic
properties of the emission mechanism. In the model with diffractive
lensing and scintillations the effect is provided by small velocities
of the lenses and the scintillating media and by the randomized
central frequency of the FRB progenitor. The FRB   signals acquire the
same narrow--band spectral structure if they are  localized in the
same frequency band  and occur shortly after one another, so that the
lens and the medium do not have enough time to evolve.

\begin{figure}
  \unitlength=1mm
  \begin{picture}(86,100)
    \put(0,0){\includegraphics[width=8.6cm]{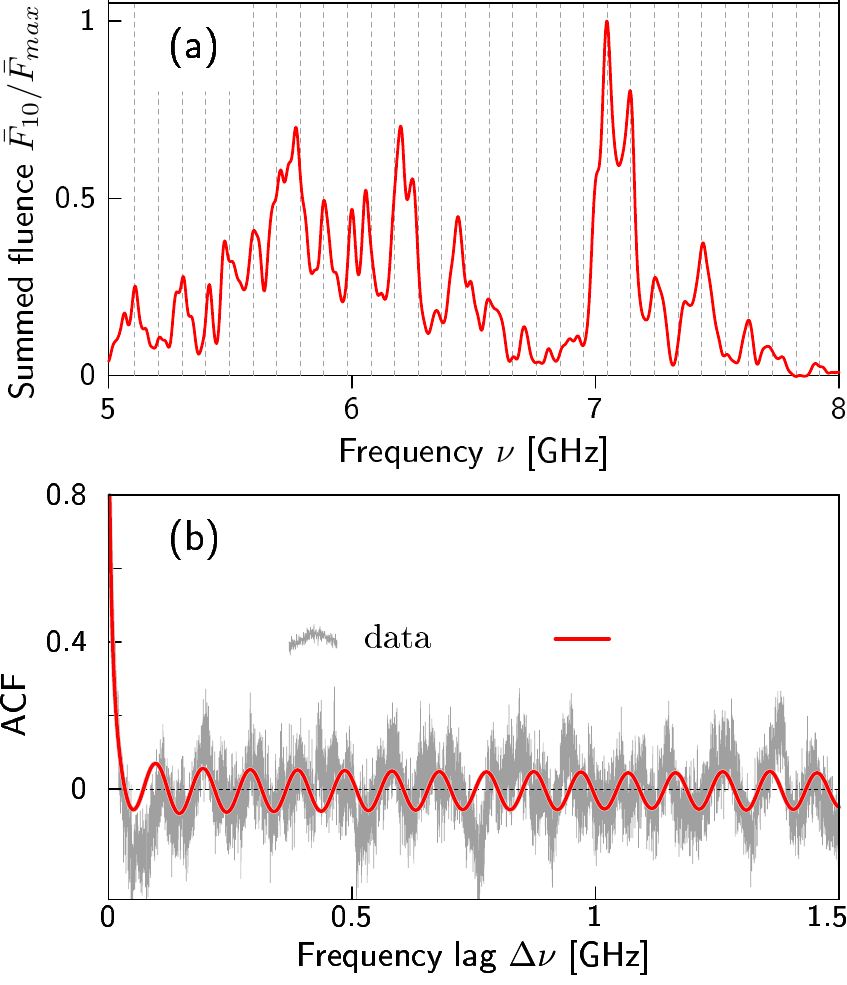}}
    \put(65,34){\fontsize{10}{1}\sffamily Eq.~(\ref{eq:59})}
  \end{picture}
\caption{(a) The sum of the normalized spectra for the 12 most
    powerful bursts. The result is smoothed with $\sigma = 10\;
  \mbox{MHz}$  and shown in units of the maximum. The lattice of
  equidistant vertical 
    lines has spacing ${T_\nu = 97 \; \mbox{GHz}}$. (b) ACF  
    of the summed spectrum and its fit with Eq.~(\ref{eq:59}).} 
\label{fig:sum}
\end{figure}

We perform another important test by summing up the normalized
  spectra of the 12 most powerful bursts. The result is more
  noisy\footnote{Another exercise is to sum up the original,
      unnormalized spectra. However, that sum is dominated by the 
      contributions of the strongest bursts 11A and 11D and the
      resulting ACF resembles Figs.~\ref{fig:ACF},
      \ref{fig:acf_11D}.} than the strongest 11D and 11A spectra 
  because the weaker bursts give the same--order contributions into
  the normalized sum.  The summed
  spectrum is visualized in Fig.~\ref{fig:ACF}, where smoothing with
  $\sigma = 10\;\mbox{MHz}$ is used. One finds that many of its maxima
  appear near the periodic lattice of dashed vertical lines. Besides, 
  the ACF of this summed spectrum (Fig.~\ref{fig:sum}b) includes many
  almost equidistant peaks at large $\Delta 
  \nu$. Fitting this function with Eq.~(\ref{eq:59}), we obtain 
  the frequency period $T_\nu \approx 97 \; \mbox{MHz}$ and
  decorrelation bandwidth ${\nu_d \approx 4.45 \; \mbox{MHz}}$ which are
  close to our previous results. Notably, the fit is pretty good: note
  that the
  positions of the ACF peaks in Fig.~\ref{fig:sum}b correlate with the
  periodic maxima of the fitting function over many periods. We
  conclude that the periodic structure is stable in time and not
  peculiar to the strongest bursts.

It is worth discussing possible theoretical models for the lens which
splits the FRB wave into two rays and creates the periodic spectral
structure. First, it may be formed by a plasma residing,
  say, in the FRB host galaxy. Consider e.g.\ the one--dimensional
Gaussian lens of  Sec.~\ref{sec:main-peak} 
with the phase shift,
\begin{equation}
  \label{eq:64}
    \Phi_l = \frac{1}{2r_{F,\, l}'^2} \left[(x - \tilde{x}')^2
    + \alpha_l' a'^2\, \mathrm{e}^{-x^2/a'^2}\right]\;.
\end{equation}
Here we equipped all the lens parameters with the primes
and ignored the trivial dependence on $y$ leaving only one 
transverse coordinate $x$. We also changed the sign in front of the
second term, so now the lens with ${\alpha_l' \propto -\delta n_e>0}$ 
describes an {\it underdensity} of free electrons. For simplicity
below we assume $\ln\alpha_l' \gg 1$ and $\tilde{x}' \sim a' \sqrt{\ln
  \alpha_l'}$~--- a strong lens relatively far away from the line of
sight.

Note that the underdensities are expected to appear 
  in the interstellar medium due to heating, e.g., by the magnetic 
  reconnections. They were often used to explain the pulsar data. For
  example, \cite{Pen:2011dz} interpreted the pulsar extreme scattering
  events (ESEs) as lensing on the Gaussian
  underdensities Eq.~(\ref{eq:64}). Another type of underdensity
  lenses in the form of corrugated plasma sheets was suggested to
  cause pulsar scintillation arcs in \cite{2018MNRAS.478..983S}. We
  will demonstrate that the lens Eq.~(\ref{eq:64}) can explain the diffractive
  peaks in our data.

\begin{figure}
  \centerline{\includegraphics{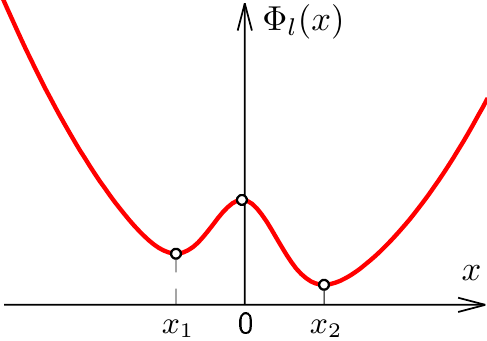}}
  \caption{Phase delay Eq.~(\ref{eq:64}) in the Gaussian underdensity lens.}
  \label{fig:phil}
\end{figure}
The phase delay Eq.~(\ref{eq:64}) is plotted in Fig.~\ref{fig:phil}. It has
three extrema corresponding to three radio rays. In
Appendix~\ref{sec:underdensity-lens} we argue that the ray with 
$x\approx 0$ has small gain factor and can be ignored, while the other
two give Eq.~(\ref{eq:63}). Computing the
parameters of these rays, we obtain 
\begin{equation}
  \label{eq:65}
  A_{\mathrm{osc}}  \approx \sqrt{1 - \frac{\tilde{x}'^2}{a'^2
      \ln \alpha_l'}} \;, \qquad
  T_\nu \approx  \frac{\pi\nu r_{F,\, l}'^2}{a' \tilde{x}' \sqrt{\ln \alpha_l'}}\;.
\end{equation}
The experimental values $A_{osc} \approx 0.5$ and $T_\nu \approx 95 \;
\mbox{MHz}$ then give ${\tilde{x} \approx  a' \sqrt{\ln
    \alpha_l'}}$ and ${r_{F,\, l}' \approx 0.07\, a'\sqrt{\ln\alpha_l'}}$,
where realistically, $\ln\alpha_l' \sim 3 - 10$.

The second option includes lensing of the FRB signal on the gravitating
compact  (point--like) object of mass $M$, e.g.\ a primordial black
hole or a dense mini--halo, like in \cite{Katz:2019qug}. The total phase delay
in  this case has the form similar to
Eq.~(\ref{eq:64}),
see~\cite{PetersonFalk}, \cite{Matsunaga:2006uc},
\cite{Bartelmann:2010fz}:
\begin{equation}
  \label{eq:74}
    \Phi_l = \frac{1}{2r_{F,\, l}'^2} \left\{ (\bm{x} - \tilde{\bm{x}}')^2
      - 2 (d_{lo}' \theta_E)^2  \ln \left| \frac{\bm{x} -
        \bm{x}_o}{d_{lo}'}\right| \right\}\,,\!\!
\end{equation}
where $d_{lo}'$ is the distance to the object, $\tilde{\bm{x}}'$ is
given by Eq.~(\ref{eq:16}) with the parameters of the new lens,
$r_{F,\, l}'$ is the respective Fresnel scale, and $\theta_E = (4G M
d_{pl}'/d_{lo}' d_{po})^{1/2}$ is the Einstein angle. Now, the second
term in the phase shift is caused by gravity rather than
refraction. That is why it  is proportional to the frequency of the
radio wave and mass of the lens: $(\theta_E/r_{F,\, l}')^2 \propto \nu
M$.

Generically, the point--like gravitational lens splits the radio wave
into two rays in Eq.~(\ref{eq:63}). Computing the respective
eikonal solutions, one finds,
\begin{align}
  \label{eq:76}
  & A_{\mathrm{osc}} = \frac{2}{\zeta^2 + 2} \;, \qquad \\
  \notag
  & T_\nu = \frac{1}{4GM} \, \left[ \zeta \sqrt{\zeta^2 + 4}  + 2 \ln
    (\zeta/2 + \sqrt{\zeta^2 /4 + 1})\right]^{-1}\;,
\end{align}
see Appendix~\ref{sec:gravitational-lens} and~\cite{PetersonFalk},
  \cite{Matsunaga:2006uc}, \cite{Bartelmann:2010fz},
  \cite{Katz:2019qug} for details. Here 
we introduced the angular separation of the source from the lens in units
of the Einstein angle ${\zeta   = |\bm{x}_p   -\bm{x}_o|/(\theta_E
  d_{lo}')}$. Substituting the mean experimental values of
$A_{\mathrm{osc}}$ and $T_\nu$, we obtain ${\zeta \approx 1.4}$ 
and $M \approx 1.1\cdot 10^{-4}\, M_{\odot}$. 

Let us guess, where the gravitational lens lives. It is natural to
  assume that such exotic objects constitute a part $\gamma$ of dark
  matter. Then the probability of meeting one of them in the
  intergalactic space at distance $|\bm{x}_p - \bm{x}_o| \lesssim
  \zeta \theta_E d_{lo}'$ from the line of sight is of order ${\gamma
  \zeta^2 G \rho_{m} d_{po}^2 \sim 10^{-2} \,\gamma}$, where we
  substituted $\zeta \approx 1.4$, the distance to the source $d_{po}\sim
  \mbox{Gpc}$, and the mean dark matter density ${\rho_m \sim 3\cdot 10^{-6}
    \; \mbox{GeV}/\mbox{cm}^3}$. Thus, the gravitational lensing of
  one FRB signal is  relatively unprobable even if all dark matter
    consists of lenses. However, the probability of the
  respective event inside the galactic halo of Mpc size is $\sim 10$
  times smaller despite the larger density. Thus, the gravitational
  lensing is generically expected to occur on the way between the galaxies.

It would be great to discriminate between the above two lenses
on the basis of the spectral data alone. For example, one may assume
that the dependence of the frequency period $T_\nu$ on frequency $\nu$ is different
in the two cases. Indeed, universality of the gravitational lensing gives
$T_\nu(\nu) = \mbox{const}$, whereas refraction of radio waves in 
the plasma is essentially $\nu$--dependent. Note, however, that the first
(geometric) term of the plasma lens phase shift Eq.~(\ref{eq:64})  is 
proportional to the frequency, just like the gravitational
shift Eq.~(\ref{eq:74}). If it is important, the dependence
of the frequency period on $\nu$ may be extremely weak.  An
example is provided above by the strong underdensity
lens. In this case the values of $A_{\mathrm{osc}}$ and $T_\nu$ in 
Eq.~(\ref{eq:65}) logarithmically depend on $\nu$ via the lens
strength $\alpha_l' \propto \nu^{-2}$ and become indistinguishable from
constants at $\alpha_l' \gtrsim 5$.

Nevertheless, it is worth stressing that the experimental data do not
indicate any dependence of the spectral oscillation parameters on
frequency. Indeed, the values of $T_\nu$ in Fig.~\ref{fig:Tnu}a are
almost the same for the bursts with essentially different central
frequencies $\nu_c$. 

\section{Comparison with earlier studies}
\label{sec:comp-with-earl}
Our analysis of the narrow--band spectral structure essentially
  differs from the previous ones. Let us explain the
    distinction and place our results in the context of the
  other FRB~20121102A studies.

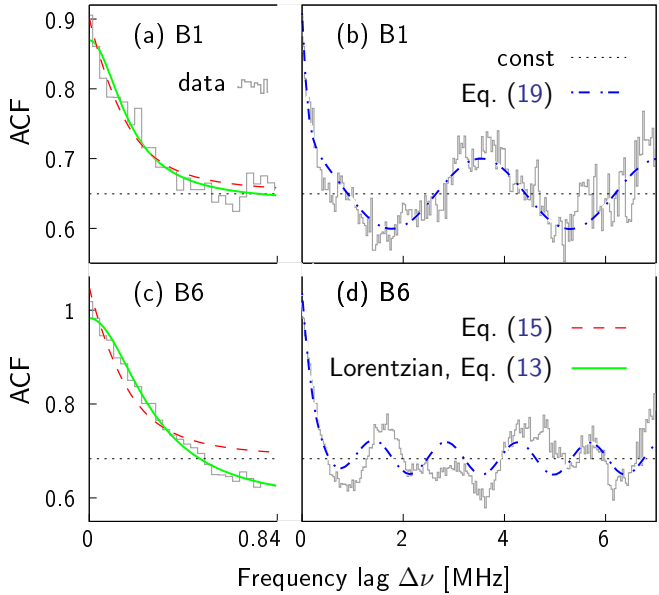
\begin{figure}
  \unitlength=1mm
  \begin{picture}(86,80)
    \put(0,0){\includegraphics[width=8.6cm]{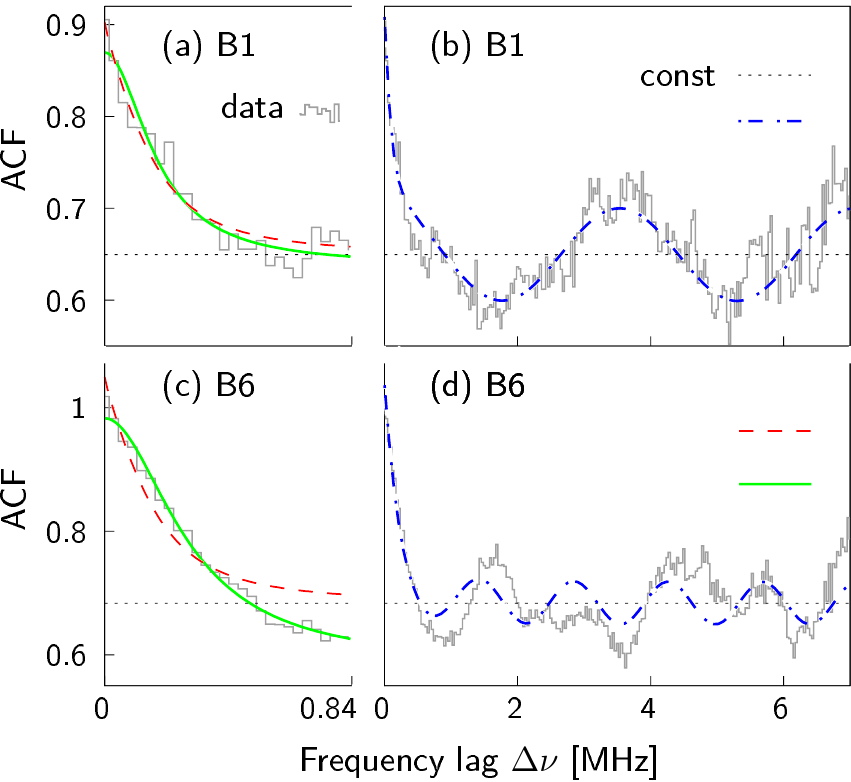}}
    \put(60,65.5){\fontsize{10}{1}\sffamily Eq.~(\ref{eq:59})}
    \put(60,34.3){\fontsize{10}{1}\sffamily Eq.~(\ref{eq:62})}
    \put(42.5,29){\fontsize{10}{1}\sffamily Lorentzian,
      Eq.~(\ref{eq:36})}
  \end{picture}
  \caption{Autocorrelation functions of the two FRB~20121102A
      bursts B1 and B6 at frequency 2.24~GHz (Figure~3
      of~\cite{Majid:2020dwq}). Panels (a) and (c) show their
      initial falloffs at ${\Delta \nu < 0.84 \;
        \mbox{MHz}}$, while (b) and (d) cover larger $\Delta \nu$
      intervals.} 
  \label{fig_ACF_DSN}
\end{figure}

The unusual features were observed in the burst autocorrelation
  functions before, but a conclusive evidence for their diffractive
  origin has never appeared. For example,  \cite{Majid:2020dwq}
  published\footnote{The data in Figure~3 of \cite{Majid:2020dwq} 
    slightly mismatch their own Lorentzian fit, so we assumed
    inaccuracies in their plot and shifted the graphs to the left by
    two bins.} two ACFs  of the FRB~20121102A bursts B1 and B6 measured
by the $\mbox{DSS--43}$ telescope of the Deep Space  Network at frequency
2.24~GHz, see Fig.~\ref{fig_ACF_DSN}. At large $\Delta \nu$  these
functions display several recognizable maxima, see
Figs.~\ref{fig_ACF_DSN}b,d. Interpreting the latter as a
manifestation of the two--wave interference, one can formally fit
the B1 and B6 ACFs with the cos--like function in the integrand of 
Eq.~(\ref{eq:59}) plus a constant. The results of this fit are
shown by the dash--dotted lines in Figs.~\ref{fig_ACF_DSN}b,d. The
respective best--fit frequency periods are $T_{\nu} \approx 3.5$
and  $1.4 \; \mbox{MHz}$ for the bursts B1 and B6, respectively.

Note, however, that unlike the BL digital backend of the
  Green Bank Telescope, the 
  instrument of~\cite{Majid:2020dwq} has narrow 8~MHz
  non--contiguous sub-bands. As a consequence, the graphs B1 and  B6 in
  Fig.~\ref{fig_ACF_DSN} include only 2 and 4 peaks in the available
  frequency interval, and one cannot judge whether they are periodic
  or not. Compare this to the strongest bursts 11A and 11D in
  Figs.~\ref{fig:ACF}a and~\ref{fig:acf_11D}, every one of which
  displays $\sim 10$ approximately equidistant ACF maxima. Besides,
  the apparent frequency periods $T_\nu$ of the bursts  
B1 and B6 mismatch by a factor of~2. The data of the previous Section
were consistent: the values $T_{\nu}$ of the 12 strongest
bursts were grouping within the 15\% interval near the central
value. Finally, the widths of the maxima in Fig.~\ref{fig_ACF_DSN} are
comparable to the entire frequency interval $\nu_2 - \nu_1 =
8\;\mbox{MHz}$. In fact, the presence of the unsuppressed random
fluctuations is  expected\footnote{Although is hard to estimate
    reliably the respective probability.} at
these scales, since the frequency integral in the ACF effectively kills
only the noise with $\nu_d \ll \nu_2 - \nu_1$, cf.\ Eq.~(\ref{eq:25}).

We conclude that the periodic structures cannot be distinguished from
the random spectral behavior in~\cite{Majid:2020dwq} data. To do
that, one needs wide--band measurements of many spectra, like the
  ones performed by the Green Bank Telescope.

Now, let us compare  our new method of studying
    the scintillations with the previous ones. \cite{Gajjar:2018bth}
  performed the original analysis of the 4--8~GHz Green Bank Telescope
  data by fitting the sub--band autocorrelation functions with the
  Gaussian profiles at ${\Delta \nu < 200 \; \mbox{MHz}}$. The
  resulting values of $\nu_d$~--- the half width at half maximum of
  the fitting function~--- were found to be consistent with $\nu_d
  \approx 24 \; \mbox{MHz}$ at 6~GHz, which is 
  6--7 times larger than our result. This huge discrepancy is related to
  the fact that the fitting interval $\Delta\nu < 200\; \mbox{MHz}$ in
  \cite{Gajjar:2018bth} includes the first equidistant ACF maximum, 
  see Fig.~\ref{fig:ACF}. This new feature is enveloped by the
  fitting function making the latter wider. On the other hand, we fitted
  only the initial part of the ACF to the left of its first
  minimum\footnote{Note also that our theoretical ACF does not
      resemble the Gaussian profile, see Fig.~\ref{fig:h}.}. 

\cite{Majid:2020dwq} obtained the decorrelation bandwidth of the
  two FRB 20121102A bursts B1 and B6 using the 2.4~GHz DSS--43
  telescope data. To this end the initial falloffs of the respective
  ACFs were fitted with the Lorentzian profile Eq.~(\ref{eq:36}) at
  $\Delta \nu < 0.84\; \mbox{MHz}$, where ${\cal N}$, $\nu_d$, and $c$
  served as the fit parameters, see
    \cite{1985ApJ...288..221C}. The result was $\nu_d \approx 180 \; 
  \mbox{kHz}$ and 280~kHz for the bursts B1 and B6, respectively. We
  replot the \cite{Majid:2020dwq} data and their Lorentzian fits in
  Fig.~\ref{fig_ACF_DSN} (steps and solid lines). Their procedure
  is different from ours in two important respects. First, we
  fix~$c$~--- the constant part of the ACF~--- with the
    subtraction procedure which effectively means that $c$ tracks the
    mean value of this function at large $\Delta \nu$,
  cf.~Eq.~(\ref{eq:25}). Indeed, if one leaves this parameter free
    in the fit, its value would essentially depend on the $\Delta
  \nu$ interval and, crudely speaking, would pick up the minimal
  value of the autocorrelation function at the interval boundary. This
  sensitivity is an artifact of the unexpected ACF  
  oscillations at large~$\Delta  \nu$, and it is not correct. One can
  fix the arbitrariness in our manner by fitting the ACF's at
  large $\Delta \nu$ with the constant $c$ (dotted horizontal lines in
  Fig.~\ref{fig_ACF_DSN}) and then performing the stable Lorentzian
  fit at small $\Delta \nu$. The fit result is $\nu_d(2.24\;
  \mbox{GHz}) \approx 155\pm 14\; \mbox{kHz}$ and $162 \pm
    20\; \mbox{kHz}$ for for the bursts B1 and B6,
  respectively. Notably, the two values of the decorrelation bandwidth
  now agree within the errorbars obtained from the fits.

Second, we use new theoretical profile Eq.~(\ref{eq:75}), (\ref{eq:62})
for the ACF which has sharper behavior as ${\Delta \nu \to 0}$, see
Fig.~\ref{fig:h}. As discussed in the previous Section, this method
generically gives 20\% smaller value of~$\nu_d$. Indeed, for the
bursts B1 and B6 we obtain $\nu_d \approx 128\; \mbox{kHz}$ and
133~kHz at a reference frequency 2.24~GHz, see the dashed lines
in Figs.~\ref{fig_ACF_DSN}a,c.

Now, we rescale the decorrelation bandwidth ${\nu_d \approx 130\;
\mbox{kHz}}$ at 2.24~GHz to our frequencies. Using the Kolmogorov
formula~(\ref{eq:54}) {with $\nu_d\propto \nu^{4.4}$}, we obtain
$\nu_d\approx 9.9\; \mbox{MHz}$  at 6~GHz,  which  is 2--3 times
  larger than our values: recall that the ``scintillations'' and
  ``scintillations+lensing'' fits of the previous Sections give
  ${\nu_d(6\; \mbox{GHz})\approx  4.3}$ and~3.3~MHz, respectively. This means
  that  non--Kolmogorov frequency dependence $\nu_d \propto
  \nu^{\alpha}$ with $\alpha < 4.4$ is favored by the data. In
  particular, rescaling with $\alpha = 4$ gives 
  $\nu_d \approx 6.7$~MHz at 6~GHz which differs from our
  values by the ``reasonable'' factor of 2. At even smaller
  $\alpha= 3.5$ one finds $\nu_d(6\; \mbox{GHz}) \approx 4.1$~MHz
  in agreement with our result.

Note that non--Kolmogorov scaling $\nu_d \propto \nu^\alpha$ with
  ${\alpha  < 4.4}$ was observed in the Milky Way pulsar
    signals, see e.g.~\cite{Bhat:2004xt}. Moreover the pulsar data
    coming from certain directions suggest $\alpha  = 3 - 4$
    which is not theoretically possible for weakly coupled
    turbulent plasma. Our result is of this kind.

Another value $\nu_d\approx 58\; \mbox{kHz}$ at 1.65~GHz was obtained
  by ~\cite{Hessels:2018mvq} using European VLBI Network data for one
  FRB 20121102A burst. This result corresponds to $\nu_d \approx 17$, 10,~and
5.3~MHz at 6~GHz in the cases $\alpha = 4.4$, $4$, and $3.5$,
respectively. The result at $\alpha = 4$ is again twice larger
  than ours while the one with $\alpha = 3.5$ is a good match.

Finally, let us compare the value of $\nu_d$ with the prediction
  of NE2001 model for the Milky Way distribution of free electrons 
  \citep{2002astro.ph..7156C}. Extracting the scattering measure in  
  the direction of FRB 20121102A from the provided software and
  using\footnote{For thin scattering screens inside the galaxy
      this equation agrees with our Eqs.~(\ref{eq:54}), (\ref{eq:9}).}
  Eq.~(10) of \cite{2002astro.ph..7156C}, we obtain the scintillation
  bandwidth in our galaxy: $\nu_{d,\; gal} \approx 21 \,
  \mbox{MHz}$ at 6~GHz. This result again assumes Kolmogorov scaling and it is
  5--6 times larger than our experimental values. To feel how the prediction of the
  model would change in the non--Kolmogorov case,   recall
  that  \cite{2002astro.ph..7156C} mostly use the pulsar data with 
  typical frequencies in the GHz
  range, cf.~\cite{Majid:2020dwq}. Thus, scaling with powers 
  $\alpha = 4$ and 3.5 would give $6^{4.4-\alpha}$ times smaller
  bandwidths ${\nu_{d,\; gal} \approx 10}$ and 4.2~MHz. Once
    again, we obtain factor two difference and a perfect match at
    $\alpha = 4$ and $3.5$, respectively.

To sum up, the bandwidth $\nu_d$ of our narrow--band
  scintillations differs by factors of several from the two measurements
  at other frequencies and from the prediction of the NE2001 model if
  the Kolmogorov scaling is assumed. In the case of non--Kolmogorov
  frequency dependence $\nu_d \propto \nu^{4}$ the four results
    agree up to a factor of two. If  $\nu_d \propto \nu^{3.5}$, all
    results match perfectly. Note that non--Kolmogorov scaling
    with $\alpha = 3.5$ or~4 does not contradict to our data in
  Figs.~\ref{fig:nud}b and~\ref{fig:nud_periodic}b (dotted lines) and
  in fact is observed in the pulsar measurements~\citep{Bhat:2004xt,
    Geyer:2017lbi}.

\section{Conclusions and Discussion}
\label{sec:conclusions}
In this paper we reanalyzed the spectra of FRB~20121102A measured
by~\cite{Gajjar:2018bth}. We developed practical theoretical tools
to study the random spectral components, regular peaks and periodic
spectral structures which may be caused by interstellar
scintillations, refractive  and  diffractive lensing, or, alternatively,
  can be intrinsic to the FRB progenitor.

We saw that the caustics of the refractive lens produce a spectral peak of a
distinctive recognizable form~\citep{Clegg:1997ya,   Cordes:2017eug}
that can be directly fitted  to the spectra. On the  other hand,
separation of diffractive lensing from scintillations requires
calculation of an integral  
observable: the spectral autocorrelation  function (ACF) in
Eq.~(\ref{eq:25}). The scintillations are responsible for the
monotonic falloff of this function with  frequency lag $\Delta \nu$,
while the two--ray diffraction introduces a distinctive oscillatory
behavior i.e.\ the series of pronounced equidistant maxima. We derived
explicit theoretical expressions for the ACFs which include the
effects of  Kolmogorov--type scintillations and scintillations on top
of diffractive lensing, Eqs.~(\ref{eq:55}), (\ref{eq:75}) and (\ref{eq:59}),
respectively. The latter expressions can be used to interpret the
experimental data, and in fact, fit them quite nicely. An alternative
data analysis may involve Fourier transform as in~\cite{Katz:2019qug},
periodogram method in~\cite{Zechmeister:2009js}, \cite{Ivanov:2018byi},
or Kolmogorov--Smirnov--Kuiper test in~\cite{NR}.

Using the above tools, we identify and explain several remarkable
features in the FRB 20121102A spectra. First and most importantly,
we discover a set of almost equidistant spectral peaks separated
by ${T_\nu = 95 \pm 16 \; \mbox{MHz}}$. This periodicity is a
benchmark property of wave diffraction, and we show that it
may be relevant, indeed. On the one hand, the peaks may be caused by 
diffractive gravitational (femto)lensing of the FRB signals on a
compact object of mass $10^{-4}\, M_{\odot}$, e.g.\ a primordial black
hole or a dense 
minihalo, see~\cite{Katz:2019qug}. Theoretically, such events are
expected to occur with relatively small probability $\sim 10^{-2}$ in
the intergalactic space if all dark matter consists of lenses. On
the other hand, the periodic peaks may originate from the diffractive
lensing on a plasma underdensity in the host galaxy. The respective
lenses~--- the holes in the electron density~--- may be expected to
appear due to plasma heating and in fact, are discussed in the
literature, see e.g.\ \cite{Pen:2011dz} and
\cite{2018MNRAS.478..983S}. 

Yet another suggestion would be to
attribute the periodic structure to the progenitor spectrum.
Notably, the banded pattern resembling our periodic structure has
been observed in the spectra of Crab pulsar, see
e.g.\ \cite{Hankins:2016bbl}. This may point at the same physical
origin of the two effects or similar propagation effects near
  the sources. During years, propagation and direct emission models
were proposed for explanation of the Crab bands, but none of them
has become universally accepted by now, see discussion in
\cite{Hankins:2016bbl}.

The second spectral feature is a strong, almost monochromatic peak at
7.1 GHz dominating the spectra of  most bursts. This peak was also
spotted by~\cite{Gajjar:2018bth}. We demonstrated that it can be
  produced by refractive lensing of the FRB wave on one--dimensional Gaussian
  plasma cloud. The latter may  represent long ionized filament from
  the supernova remnant in the host galaxy~\citep{Michilli:2018zec}
  or an  AU--sized elongated turbulent overdensity which are expected
  to be   abundant in galactic plasmas, see~\cite{1987Natur.326..675F, 
    Bannister:2016zqe,  Coles:2015uia}. Note also that the origin
  of the lens may be essentially different. For example, FRB
  20180916B~--- a repeating source 
  very similar to FRB 20121102A~--- is possibly a high--mass X--ray
  binary system that includes a neutron star interacting with the
  ionized wind of the companion \citep{Tendulkar:2020npy,
    Pleunis:2020vug}. In this case extreme plasma lensing may occur on
  the wind, cf.\ \cite{Main:2018kfc}.

One can still imagine that the agreement of the 7.1~GHz peak
profile with the expected spectrum of the lens is a coincidence and
this feature belongs to the intrinsic spectrum of the FRB
progenitor. Going to the extreme, one can even assume
that the progenitor produces a single line of  powerful
monochromatic emission at $7.1\; \mbox{GHz}$, and all other
frequencies add up afterwards in the course of nonlinear wave
propagation through the surrounding plasma. The generation
mechanisms for the monochromatic signals use cosmic masers
\citep{Lu:2017prv} or Bose stars made of dark matter 
axions~\citep{Tkachev:1986tr} that decay into photons. The latter
process may occur explosively in strong magnetic 
fields~\citep{Iwazaki:2014wka, Tkachev:2014dpa, Pshirkov:2016bjr} or in
the situation of parametric resonance~\citep{Tkachev:1986tr,
  Tkachev:2014dpa, Hertzberg:2018zte, Levkov:2020txo,
  Hertzberg:2020dbk, Amin:2020vja}. However, the axion--related
mechanisms still belong 
to the speculative part of the FRB theory, whereas the cosmic masers with
realistic parameters fail to provide the required FRB
luminosity, see~\cite{Lu:2017prv}.

Third, the FRB signals illuminate parts of a global GHz--scale
spectral structure, cf.~\cite{Sobacchi:2020pbe}. The imprint of
  this structure was previously noticed
by~\cite{Gajjar:2018bth} in the summed 4--8 GHz spectrum of FRB
20121102A. We demonstrate that the structure drifts linearly
with time and therefore presumably represents a propagation
effect e.g.\ GHz--scale scintillations. Of course, even this last
feature may belong to the FRB source if the emission region
  itself evolves linearly.

Fourth, all pieces of the propagation scenario fit together if 
the spectrum of the FRB  20121102A progenitor has a relatively
  narrow bandwidth $\nu_{bw} \sim \mbox{GHz}$, and its central
  frequency changes rapidly and significantly from burst to
  burst. The same properties were 
  observed before in the registered spectra of FRB 20121102A
  \citep{2017ApJ...850...76L,  2019ApJ...877L..19G, Gajjar:2018bth,
    Hessels:2018mvq, Majid:2020dwq} and FRB 20180916B
  \citep{Chawla:2020rds, Pearlman:2020sox}, but in these studies the
  intrinsic progenitor properties were not separated from the
  propagation phenomena. We perform the separation and in fact, use
  the   propagation effects as landmarks for studying the progenitor 
  spectrum.

In particular, the ``main''peak at 7.1~GHz, which we  attribute to
  the strong lens, disappears  
in some spectra reappearing in  later--coming
bursts at the same position and  with the same form. We explain that
this happens precisely because the progenitor spectrum has 
 a  GHz bandwidth and variable central frequency. Indeed, all the spectra
  with the main peak are located within the GHz band around 7.1~GHz,
  and the spectra without  it have the major power outside of this
  band. Further, the bursts 
happening at different times illuminate different parts of the linearly
evolving wide--band spectral pattern introduced  above. Reconstructing
the pattern, we estimate the bandwidth of the progenitor.

Finally, we develop new generalized framework for the
  analysis of strong interstellar scintillations. We obtain
  their  decorrelation bandwidth $\nu_d$ by fitting the experimental
  ACFs 
  with the new theoretically derived profile.  Notably, the ACF data
  agree with the theory, the value of $\nu_d$ crudely respects the
  predicted power--law scaling, and the overall scintillation  pattern
  slowly drifts in frequency due to motion of the observer relative to
  the scintillating clouds. Our result for $\nu_d$ slightly
  depends on the assumptions on the above--mentioned periodic
  spectral structure. If we ignore it and use the scintillations--only model,
  the best--fit value is ${\nu_d = 4.3\pm 0.9 \; \mbox{MHz}}$ at the
  reference frequency 6~GHz. Adding the periodic structure to the fit,
  we obtain a consistent value ${\nu_d (6 \; \mbox{GHz}) = 3.3\pm 0.6
    \; \mbox{MHz}}$. One can conservatively consider the difference
between the two results as a systematic error, although we do suggest
that the last result is more consistent. We believe that our method
for extracting $\nu_d$ is more reliable than the
previous ones because it uses the theoretically predicted ACF profile.

Note that the  narrow--band scintillations were observed in the
  FRB~20121102A spectra  before~by \cite{Hessels:2018mvq} at
  $1.65~\mbox{GHz}$ and \cite{Majid:2020dwq} at
  $2.24~\mbox{GHz}$. We perform comparison by scaling their two
  values of $\nu_d$ to 6~GHz with the power law $\nu_d(\nu)
  \propto \nu^\alpha$. In the  Kolmogorov case $\alpha = 4.4$
  the scaling gives 5 and 3 times larger bandwidths than our
    result, respectively. Thus, the data strongly favor non--Kolmogorov
  frequency dependence. The smallest theoretically motivated power for
  the weak turbulence is $\alpha = 4$. In this case the values of
  \cite{Hessels:2018mvq} and \cite{Majid:2020dwq}  at 6~GHz
  differ from ours by the factors of 3 and 2, respectively, which is already
  tolerable given large experimental uncertainties and different
  instruments. If $\alpha = 3.5$, the three experimental results
  agree.

Do the scintillations appear in the Milky Way or in the FRB host
  galaxy? The model NE2001 \citep{2002astro.ph..7156C} predicts
  Milky Way scintillations with $\nu_{d,\; gal}(6\; \mbox{GHz}) \approx
  21\; \mbox{MHz}$ in the direction of FRB 20121102A,  and that is
    6 times larger than our result. But the same model assumes
  Kolmogorov scaling of $\nu_d$ with $\alpha = 4.4$. Thus, the
  discrepancy can be again attributed to deviations from this
  law. Indeed, following \cite{Majid:2020dwq} we crudely account for
  arbitrary $\alpha$ in the model and arrive to estimates ${\nu_d(6\;
  \mbox{GHz}) \sim 10\; \mbox{MHz}}$ and $4\; \mbox{MHz}$ at $\alpha =
  4$ and $3.5$, which are closer to our result. Thus, the
  scintillations presumably originate in our Galaxy.

To conclude,  the studies of the FRB spectra are still in their infancy,
but they evolve fast. With clever data analysis and
  separation of propagation effects, they soon may be able to purify the
pristine chaos of the present--day theory for the FRB engines down to
a single graceful picture.

\acknowledgments{
  We thank V.~Gajjar for help, S.~Sibiryakov for discussions, and
  the Referee for criticism. Scintillations in the FRB~20121102A
  spectra were studied within the framework of the RSF grant
  16-12-10494. Investigation of the FRB lensing was supported by the
  Ministry of Science and Higher Education of the Russian Federation
  under the contract 075-15-2020-778 (State project “Science”). The
  rest  of this paper was funded by the Foundation for the Advancement
  of Theoretical Physics and Mathematics ``BASIS.'' Numerical
  calculations were performed on the Computational cluster of Theory 
  Division of~INR~RAS.} 

\appendix

\twocolumngrid

\section{Computing the spectra}
\label{sec:computing-spectra}
Let us explain the computation of the spectral fluence Eq.~(\ref{eq:1}) in
detail. The main idea is to choose the signal region $t_1(\nu) < t <
t_2(\nu)$ which minimizes the instrumental noise.

\begin{figure}
  \centerline{\includegraphics{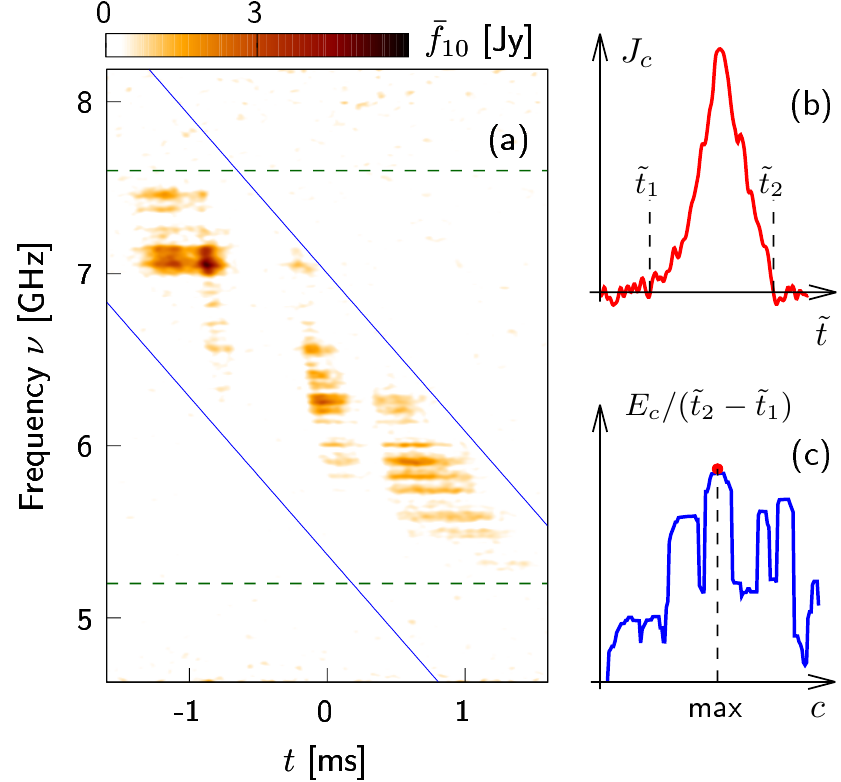}}
  \caption{Computing the signal region for the burst 11A.
    \label{fig:cutoff}}
\end{figure}

Most of the bursts are tilted in the $t-\nu$ plane, even after
de--dispersion described in~\cite{Gajjar:2018bth}, see the burst 11A in
Fig.~\ref{fig:cutoff}. We therefore use the linear  boundaries   ${t_1(\nu)
= \tilde{t}_1 + c\nu}$ and $t_2(\nu) = \tilde{t}_2 + c\nu$ of the
signal region with frequency--independent signal duration $t_2 -
t_1$. 

To determine the parameters $\tilde{t}_1$, $\tilde{t}_2$, and $c$, we
employ two auxiliary technical steps. First, we Gauss-average the signal
$f(t,\, \nu)$ over the moving time and frequency windows $\sigma_t =
0.02\; \mbox{ms}$ and $\sigma = 10\; \mbox{MHz}$,
cf.\ Eq.~(\ref{eq:2}). Second, for every burst we preselect the 
frequency interval $\nu_1 < \nu < \nu_2$ including the signal (dashed
lines in Fig.~\ref{fig:cutoff}).

Once this is done, we integrate the smoothed density $\bar{f}_{10}$ along
the inclined line,
\begin{equation}
  \notag
  J_c(\tilde{t}) = \int\limits_{\nu_1}^{\nu_2} d\nu \, \bar{f}_{10}(\tilde{t}
  + c \nu,\, \nu)\;.
\end{equation}
This function is positive if the line $t = \tilde{t} + c\nu$ crosses
the signal. Outside of the signal region
$J_c(\tilde{t})$ oscillates near zero due to noise. We
therefore select $\tilde{t}_1$ and $\tilde{t}_2$ to be the first 
zeros of this function surrounding its global maximum, see
Fig.~\ref{fig:cutoff}b. 

To choose the optimal value of the tilt $c$, we note that the integral
$E_c = \int_{\tilde{t}_1}^{\tilde{t}_2} J_c(\tilde{t}) \,
d\tilde{t}$ estimates the energy within the signal region. We
maximize the average signal power ${E_c/(\tilde{t}_2 - \tilde{t}_1)}$ with
respect to $c$ thus choosing the minimal region $t_1 < t < t_2$ with 
major part of the total energy, see Fig.~\ref{fig:cutoff}c.

Once the signal region is identified, we perform integrations in
Eqs.~(\ref{eq:2}), (\ref{eq:1}) obtaining\footnote{The burst 12B
  includes two well separated parts. In this case we use two signal
  regions with different tilts~$c$.} the spectral fluence
$\bar{F}_{10}(\nu)$. Note that this quantity is computed at all
frequencies, even outside of the auxiliary interval $\nu_1 < \nu <
\nu_2$. At $\nu < \nu_1$ and $\nu > \nu_2$ the spectral fluence
fluctuates near zero due to noise. 

The experimental errors are estimated assuming that the statistical
properties of the measuring device are time--independent. For every
frequency $\nu$ we select a sufficient number of random time points
outside of the signal region, combine the data at these points into an
artificial interval $t_1(\nu) < t < t_2(\nu)$, then use
Eqs.~(\ref{eq:2}), (\ref{eq:1}). This gives the random fluence $\Delta
\bar{F}_{10}(\nu)$ of  the noise. We finally subtract the statistical
mean of $\Delta \bar{F}_{10}$ from Eq.~(\ref{eq:1}) and use its 
standard deviation to estimate the errors (shaded areas in
Fig.~\ref{fig:spectrumAC}). 

\section{Lensing}
\label{sec:lenses}

\subsection{Eikonal approximation}
\label{sec:generalities}
In this Appendix we review the effects of plasma and gravitational
lenses on the propagating FRB signals. A plasma cloud equips the
  radio wave with a dispersive phase shift $\varphi_l$,
\begin{equation}
  \label{eq:17}
  \varphi_l = -\frac{r_e}{\nu} \, \mbox{DM}_l(\bm{x}) =
  -\frac{r_e}{\nu}  \int dz \; \delta n_e({\bm x},\, z)\;,
\end{equation}
where $\delta n_e$ is the overdensity of free electrons, $r_e\equiv
e^2 / m_e$ is the classical electron radius, and the integral runs
along the line of sight. Notably, the dispersive phase Eq.~(\ref{eq:17})
is inversely proportional to frequency.

We will use the standard~\citep{Rickett1990, Katz:2019qug} assumption
that the entire dispersive
shift Eq.~(\ref{eq:17}) is acquired on the relatively thin 
``lens screen'' halfway through the plasma cloud, see the dashed line
in  Fig.~\ref{fig:lens}. This approach explicitly separates the geometric
and dispersive effects. It is justified if the lens is spatially
separated from the other propagation phenomena.

Now, the dispersive phase Eq.~(\ref{eq:17}) is a function of the
two--coordinate ${\bm x} = (x,\, y)$ on the lens screen. Besides, the
radio ray $p{\bm x}o$ in Fig.~\ref{fig:lens} consists of two straight
parts giving the extra geometric phase shift
\begin{equation}
  \label{eq:18}
  \phi_l = 2\pi \nu (r_{p{\bm x}} + r_{{\bm x} o} - r_{po}) \approx
  ({\bm x} - \tilde{\bm x})^2 / 2r_{F, \, l}^2\;.
\end{equation}
Here $r_{p{\bm x}}$, $r_{{\bm x} o}$, and $r_{po}$ are the distances
between the respective points in Fig.~\ref{fig:lens}; in the last
equality we used the small--angle approximation $r_{ij} \approx d_{ij}
+ ({\bm x}_i - {\bm x}_j)^2/2d_{ij}$ and collected the total square.
Recall that ${\bm x}_p$ and ${\bm x}_o$ are the progenitor and observer
shifts, $\tilde{\bm{x}}$  is defined in Eq.~(\ref{eq:16}), and
$r_{F,\, l}$ is the lens Fresnel scale, see Sec.~\ref{sec:main-peak}. The
total phase shift 
\begin{equation}
  \label{eq:68}
  \Phi_l(\bm{x}) = \phi_l(\bm{x}) + \varphi_l(\bm{x})
\end{equation}
is the sum of Eqs.~(\ref{eq:18}) and (\ref{eq:17}). 

The gravitational lens modifies the phases of radio waves is a
different way: by gravitationally attracting the radio rays and
changing their length. For convenience we will also divide its phase 
shift  $\Phi_l$ into the ``naive'' geometric part Eq.~\eqref{eq:18} and a
correction  $\varphi_l(\bm x)$  at the ``lens screen.'' Both $\phi_l$
and $\varphi_l$ in this case are proportional to the frequency, in
contrast to $\nu^{-1}$ behavior of  the dispersive shift in
Eq.~(\ref{eq:17}).

The complex amplitude of the observed FRB signal is given by the
Fresnel integral
\begin{equation}
  \label{eq:19}
  f_\nu = f_p\int \frac{d^2 {\bm x}}{2 \pi i r_{F,\, l}^2}  \;\;
  \mathrm{e}^{i\Phi_l ({\bm x}) }\;,
\end{equation}
where $f_p(\nu)$ is the signal of the FRB progenitor.

In this paper we consider lensing in the limit of geometric optics
$|{\bm x}| \gg r_{F,\,  l}$ when the integral (\ref{eq:19})
receives main contributions near the stationary points ${\bm x} = {\bm
  x}_j$ of the total phase. These points  satisfy the equation, 
\begin{equation}
  \label{eq:20}
  {\bm \nabla}_{\bm x} \Phi_l({\bm x}) \equiv r_{F,\, l}^{-2} ({\bm x} - \tilde{\bm
    x}) + {\bm \nabla}_{\bm x} \varphi_l({\bm x}) = 0\;.
\end{equation}
The signal~(\ref{eq:19}) is then given by the saddle--point formula, 
\begin{equation}
  \label{eq:21}
  f_\nu = f_p \sum\limits_j\; G_{j}^{1/2} \;
  \mathrm{e}^{i\Phi_l({\bm x}_j)}\;,
\end{equation}
where we introduced the gain factors of the radio paths, 
\begin{equation}
  \label{eq:69}
  G_j(\nu) = \Big|\det(\delta_{\alpha\beta}
  + r_{F,\, l}^2 \partial_\alpha \partial_\beta
  \varphi_l)\Big|_{\bm{x} = \bm{x}_j}^{-1} 
\end{equation}
with ${\partial_\alpha \equiv \partial / \partial x_\alpha}$. Note
that the determinant in Eq.~(\ref{eq:69}) is not necessarily
positive. For convenience we keep $G_j >0$ and include the
phase of the determinant into  ${\Phi_l(\bm{x}_j) \to \Phi_l(\bm{x}_j)
  - \pi 
  n_j/2}$, where $n_j = 0,1,2$ if $\bm{x} = \bm{x}_j$ is a minimum,
a saddle point, and a maximum of $\Phi_l$, respectively. Once $f_\nu$ is
computed, one obtains the fluence $F(\nu) \equiv |f_\nu|^2$.

\subsection{Refractive and diffractive lenses}
\label{sec:refr-diffr-lens}
We see that every radio path ${\bm x}_j$ adds the term
${\Delta F = G_j |f_p|^2}$ to the fluence, while its interference with
other paths produces oscillating terms proportional to
$\cos(\Phi_j - \Phi_k)$, where $\Phi_j \equiv \Phi_l({\bm x}_j)$. For
example, in the case of two trajectories Eq.~(\ref{eq:21}) gives,
\begin{equation}
  \label{eq:22}
  F = F_{p} 
    \left[G_1 + G_2 + 2 (G_{1}G_2)^{1/2} \; \cos(\Phi_1 -
    \Phi_2)\right]\;,
\end{equation}
with $F \equiv |f_\nu|^2$ and $F_{p} \equiv |f_p|^2$ denoting the
registered and source fluences, respectively.
The last term in Eq.~(\ref{eq:22}) represents diffraction. As a
function of frequency, it oscillates with the period
\begin{equation}
  \label{eq:23}
  T_\nu = 2\pi |\partial_\nu (\Phi_1 - \Phi_2)|^{-1} \sim O(2 \pi \nu \,
  r_{F,\, l}^2/a^2)\;,
\end{equation}
where $a \sim |{\bm x}_2 - \bm{x}_1|$ is the typical lens size.

In Sec~\ref{sec:main-peak} we consider a refractive lens with extremely
large ${a/r_{F,\,  l} \gg (\nu/\sigma)^{1/2}}$. In this case the
oscillatory term in Eq.~(\ref{eq:22}) is exponentially dumped by the
instrumental smoothing Eq.~(\ref{eq:2}) with window $\sigma$. As a
  consequence, the main effect of this lens is to multiply the source
fluence with the sum of gain factors in Eq.~(\ref{eq:11}).

In Sec.~\ref{sec:periodic-structure} we interpret the periodic spectral
structures with $T_{\nu} \sim 100\; \mbox{MHz}$ as the oscillating 
term in Eq.~(\ref{eq:22}). Of course, all these structures may be
killed by smoothing with sufficiently large window, say, $\sigma =
50\; \mbox{MHz}$. In this case $\bar{F}_{50} \approx F_p (G_1 + G_2)$. The
theoretical lens signal Eq.~(\ref{eq:22}) then can be rewritten as
\begin{equation}
  \label{eq:24}
  F \approx \bar{F}_{50} \left[ 1+ A_{\mathrm{osc}} \cos(\Phi_1 - \Phi_2)\right]\;,
\end{equation}
where
\begin{equation}
  \label{eq:73}
  A_{\mathrm{osc}} = 2 (G_1 G_2)^{1/2} \; (G_1 +
  G_2)^{-1} \leq 1
\end{equation}
is the relative oscillation amplitude.  In the main text we also compute the 
correlation function Eq.~\eqref{eq:25} by multiplying $\delta F(\nu) = F -
\bar{F}_{50}$ at closeby frequencies $\nu$ and $\nu + \Delta \nu$ and
integrating over~$\nu$. This procedure exponentially suppresses the
terms oscillating with the integration variable $\nu$ leaving
\begin{multline}
  \label{eq:26}
  \mbox{ACF}(\Delta \nu)  =  \frac{\cal N}{2}
        \int\limits_{\nu_1}^{\nu_2 - \Delta \nu} d\nu \;\; \frac{\bar{F}_{50}(\nu)
            \bar{F}_{50}(\nu+\Delta \nu) 
        }{\nu_2 - \nu_1  - \Delta \nu} \\[.5em] \times
          A_{\mathrm{osc}}(\nu)\, A_{\mathrm{osc}}(\nu + \Delta
        \nu) \,  \cos(2\pi \Delta \nu/T_\nu)\;.
\end{multline}
In the case of narrow--bandwidth spectra with ${\nu_2 - \nu_1 \ll
  \nu}$ and $\Delta \nu \ll \nu$  one can ignore the frequency
dependence the period and find,
\begin{equation}
  \label{eq:27}
  \mbox{ACF} \approx \cos(2\pi \Delta \nu/T_\nu)\;,
\end{equation}
where the normalization was performed, $\mbox{ACF}(0) = 1$. In practice,
Eq.~(\ref{eq:26}) accounts for the wide--band envelope $\bar{F}_{50}$
of the spectrum and therefore better fits the experimental data,
though Eq.~(\ref{eq:27}) is simpler and may be used on preparatory 
stages.

\subsection{Gaussian overdensity lens}
\label{sec:lens-caustics}
In Sec.~\ref{sec:main-peak} we consider a Gaussian lens with the
profile ${\varphi_l = -(r_e/\nu) \, \mbox{DM}_l\,
  \mathrm{e}^{-x^2/a^2}}$  depending only on one transverse coordinate
$x$. The total phase shift Eqs.~(\ref{eq:18}), (\ref{eq:68}) in this case
reduces to Eq.~(\ref{eq:3}). The $y$ component of the eikonal equation
(\ref{eq:20}) gives $y_j = \tilde{y}$ implying that the lens
bends the radio waves only in the $x$ direction. The other, $x$
component, has the form
\begin{equation}
  \label{eq:28}
  f(u) \equiv  u - \tilde{u} + \alpha u \mathrm{e}^{-u^2}  = 0\;,
\end{equation}
where $u = x/a$ and $\tilde{u} = \tilde{x}/a$. We denote the
solutions of this equation by $u_j$. The net gain factor Eq.~(\ref{eq:11})
equals $G(\nu) = \sum_j |\partial_u f(u_j)|^{-1}$.

Let us show that the lens caustics~--- solutions $u_*$ of
Eq.~(\ref{eq:28}) with infinite gain factor $G$~--- exist only if
$\tilde{x}/a = \tilde{u}$ exceeds the critical value
$\tilde{u}_{\mathrm{cr}} =  (3/2)^{3/2}$. Indeed, by definition $u_*$
satisfy equations $f(u_*) = {\partial_u f(u_{*}) = 0}$ which
can be written in the form ${\alpha_l = \mathrm{e}^{u_*^2} / (2u_*^2 -
1)}$ and $\tilde{u} =  2 u_*^3 / (2u_*^2 - 1)$. The right--hand side of
the last equation is  bounded from below by the global minimum
$\tilde{u} \geq \tilde{u}_{\mathrm{cr}}$ which occurs at $u_* =
u_{\mathrm{cr}} \equiv \sqrt{3/2}$ and $\alpha_l = \alpha_{\mathrm{cr}}
\equiv \frac12 \mathrm{e}^{3/2}$. We conclude that the lens caustics
exist only for overcritical lens shifts, $\tilde{u} \geq
\tilde{u}_{\mathrm{cr}}$. At $\tilde{u} = \tilde{u}_{\mathrm{cr}}$
they appear at the critical ``frequency'' $\alpha_l^{-1/2} =
\alpha_{\mathrm{cr}}^{-1/2}$ and move apart as $\tilde{u}$ 
grows.

Consider the near--critical situation when $\tilde{u}$ slightly
exceeds $\tilde{u}_{\mathrm{cr}}$. This corresponds to a nearby pair
of  caustics in the lens spectrum with $\alpha_{*\pm}$ and $u_{*\pm}$
close to $\alpha_{\mathrm{cr}}$ and $u_{\mathrm{cr}}$. Performing
the Taylor series expansion in $u - u_{\mathrm{cr}}$ and $\alpha_l -
\alpha_{\mathrm{cr}}$, we rewrite the lens equation as
\begin{multline}
  \label{eq:8}
  f \approx \tilde{u}_{\mathrm{cr}} - \tilde{u} -
  (\alpha_l/\alpha_{\mathrm{cr}} - 1) (u - 3u_{\mathrm{cr}}/2)  \\+ (u -
  u_{\mathrm{cr}})^3 = 0\;.
\end{multline}
Now, we can explicitly solve the caustic equations $f(u_{*})  =
\partial_u f(u_*) = 0$ with respect to $u_*$ and ``frequency''
$\alpha_l$ finding
\begin{align}
& u_{*\, \pm} \approx u_{\mathrm{cr}} \pm
\frac1{\sqrt{3}} (\alpha_{*\, \pm}/\alpha_{\mathrm{cr}} - 1)^{1/2}\;, \\
\label{eq:29}
& \alpha_{*\, {\pm}} \approx \alpha_{\mathrm{cr}} \left[ 3 
  \tilde{u}/\tilde{u}_{\mathrm{cr}} - 2  \pm \frac{4}{u_{\mathrm{cr}}}
   \, \left(\tilde{u}/\tilde{u}_{\mathrm{cr}}
  - 1\right)^{3/2} \right]\;,
\end{align}
where expansion in $\tilde{u} - \tilde{u}_{\mathrm{cr}}$ was 
performed, again. Since $\alpha_l \propto \nu^{-2}$, the last expression
fixes the positions of caustics $\nu_{0}\pm \Delta \nu$ of the lens
spectrum: $\alpha_l(\nu_0) = (\alpha_{*\, +} + \alpha_{*\, -})/2$ and
$\Delta \nu / \nu_0 = |\alpha_{*\, +} - \alpha_{*\,
  -}|/2\alpha_{\mathrm{cr}}$. We
obtain, 
\begin{align}
  \label{eq:6}
    & \tilde{u} \approx \tilde{u}_{\mathrm{cr}} + \tilde{u}_{\mathrm{cr}}
    \left(\frac{\Delta \nu \sqrt{3}}{4\nu_0 \sqrt{2}} \right)^{2/3} \;,\\
    \label{eq:12}
  & \alpha_l(\nu_0) \approx \alpha_{\mathrm{cr}} \left(3\tilde{u}/\tilde{u}_{\mathrm{cr}}
  - 2\right) \;.
\end{align}
In the main text these expressions are used to compute $\tilde{u} =
\tilde{x}/a$ and $\beta$. 

Now, suppose the lens shift $\tilde{u}$ is slightly below
$\tilde{u}_{\mathrm{cr}}$. In this case Eq.~(\ref{eq:8}) has only one
solution $u = u_1$, and the gain factor $G = [3(u_1 - u_{\mathrm{cr}})^2
  + 1 -   \alpha_l/\alpha_{\mathrm{cr}}]^{-1}$  is smooth. Nevertheless,
$G(u_1)$ has a sharp maximum at $u_1 = u_{\mathrm{cr}}$. Half--height
of the maximum is reached at $u_1 - u_{\mathrm{cr}} = \pm
\frac{1}{\sqrt{3}} (1 - \alpha_l /\alpha_{\mathrm{cr}})^{1/2}$. Substituting
these points into Eq.~(\ref{eq:8}), we find the ``frequency'' of the
maximum $\alpha_l = \alpha_l(\nu_0)$ and its half--height width ${\Delta
  \nu' / \nu_0 =    \Delta \alpha_l/ 2\alpha_{\mathrm{cr}}}$, 
\begin{align}
  & \tilde{u} \approx \tilde{u}_{\mathrm{cr}} - \tilde{u}_{\mathrm{cr}}
  \left(\frac{\Delta \nu' \sqrt{3}}{8\nu_0 
    \sqrt{2}} \right)^{2/3} \;,
  \label{eq:13}\\
\label{eq:14}
  & \alpha_l(\nu_0) \approx \alpha_{\mathrm{cr}}\left( 3 \tilde{u}/
  \tilde{u}_{\mathrm{cr}} - 2 \right)\;.
\end{align}
These equations relate the ``main'' spectral peak to the parameters of
the lens with $\tilde{x} < \tilde{x}_{\mathrm{cr}}$.

It is worth reminding that the analytic treatment of this Appendix is
applicable for narrow lens spectra, $\Delta \nu / \nu_0 \ll 1$. 
Notably, in this case the two-- or one--peaked lens contributions are
easily recognizable on the experimental graphs. 

\subsection{Diffractive underdensity lens}
\label{sec:underdensity-lens}

In Sec.~\ref{sec:periodic-structure} we study diffraction of radio rays splitted by the plasma lens. 
We use the same Gaussian profile of the dispersive phase shift $\varphi_l$ as before, but
with different sign in front of it, see Eq.~\eqref{eq:64}. This  lens
describes a hole in the interstellar plasma, i.e.\ an underdensity of
free electrons: ${\alpha_l'   \propto -\delta n >0}$.

The lens Eq.~\eqref{eq:64} has too many parameters and easily fits the 
experimental data, so in the main text we voluntarily choose the
simplest and most illustrative regime: strong lens relatively far away
from the line of sight, $\ln \alpha_l' \gg 1$ and $\tilde{x}'/a' \sim
\sqrt{\ln \alpha_l'}$. Then the eikonal equation~(\ref{eq:20}), 
\eqref{eq:64} gives three rays, cf.\ Fig.~\ref{fig:phil},
\begin{equation}
  \label{eq:70}
  x_{1,2} \approx \pm a'\sqrt{ \ln \alpha_l'} 
  \;, \qquad x_3 \approx -\tilde{x}/\alpha_l' \;,
\end{equation}
where corrections to $x_{1,2}$ are suppressed by $(\ln \alpha_l')^{-1}$.
Recall that the one--dimensional lens does not bend the rays in the $y$
direction: $y_j = \tilde{y}'$. Plugging the eikonal equation into
Eq.~\eqref{eq:69}, we simplify the expression for the lens gain  factor:
$G = |2x^2 / a'^2 - 2 x \tilde{x}'/a'^2 + \tilde{x}'/x|^{-1}$. Three
solutions~(\ref{eq:70}) then give,
\begin{equation}
  \label{eq:71}
  G_{1,2} \approx \frac{1}{2\ln\alpha_l'} \left( 1 \mp
  \frac{\tilde{x}'}{a' \sqrt{\ln\alpha_l'}}\right)^{-1}
\;,\;\;  G_{3} \approx \frac1{\alpha_l'}\;. 
\end{equation}
Notably, in our regime $\ln\alpha_l' \gg 1$ the contribution of the third
radio path can be ignored and we obtain the two--wave interference in
Eq.~\eqref{eq:63}. 

Computing the phases Eq.~\eqref{eq:64} of the two remaining solutions, we
obtain,
\begin{equation}
  \label{eq:72}
  \Phi_2 - \Phi_1 \approx 2a' \tilde{x}' \sqrt{\ln\alpha_l'} /r_{F,\,
    l}'^2\;.
\end{equation}
Now, the expressions~\eqref{eq:23} and~\eqref{eq:73} give the
frequency period $T_\nu$ and relative amplitude $A_{\mathrm{osc}}$ of the
interference oscillations. We  
derived Eqs.~\eqref{eq:65} from the main text.

\subsection{Gravitational lens}
\label{sec:gravitational-lens}
In the main text we speculate that the periodic spectral structure may
be explained by gravitational lensing of the FRB signals on a compact
object hiding at distance $d_{lo}'$ from us. A phase shift of the
radio waves in
the gravitational field of a point--like lens  is given by
Eq.~(\ref{eq:74}),  see also~\cite{PetersonFalk}, \cite{Matsunaga:2006uc},
  \cite{Bartelmann:2010fz}, \cite{Katz:2019qug}. The latter
    expression still has the form~\eqref{eq:68}, \eqref{eq:18}, like
  in the case of refraction, 
but with a specific $\varphi_l$  term. We treat the gravitational lens in
the same eikonal approximation as before.

The lens equation~\eqref{eq:20} has two solutions, 
\begin{equation}
  \label{eq:77}
  \bm{x}_{1,\, 2} = \bm{x}_o + \frac12\, \theta_E \, d_{lo}' \, \bm{\zeta} \left( 1 \pm
  \sqrt{1 +4\bm{\zeta}^{-2}}\right)\;,
\end{equation}
where we introduced the vector $\bm{\zeta} = (\bm{x}_p -
\bm{x}_o)/(d_{po} \theta_E)$ characterizing the angular shift of the
source from the lens in units of $\theta_E$. Equations~\eqref{eq:69} 
and~\eqref{eq:74} give,
\begin{align}
  \notag
  &G_{1,2} = \frac{\zeta^2+2}{2\zeta \sqrt{\zeta^2 + 4}} \pm \frac12
  \;,\\
  \notag
  &\Phi_{2} - \Phi_1 = \Omega \left[ \zeta \sqrt{\zeta^2+ 4} + 2\ln (\zeta/2 +
  \sqrt{\zeta^2/4 + 1})\right] - \frac{\pi}{2},
\end{align}
where $\zeta\equiv |\bm{\zeta}|$ an  we denoted $\Omega = 8\pi \nu
GM$. Using finally Eqs.~\eqref{eq:73} and 
\eqref{eq:23}, we obtain the parameters of spectral
oscillations in Eq.~\eqref{eq:76} of the main text.

\section{Scintillations $+$ lensing }
\label{sec:scintillations-1}

\subsection{Adding the scintillation screen}
\label{sec:adding-scint-scre}

In practice regular structures coexist in the FRB spectra with 
random scintillations caused by refraction of radio waves in the
turbulent interstellar clouds. To describe the latter effect theoretically,
we add  a thin transverse ``scintillation'' screen that equips any
propagating wave with a random phase $\varphi_S({\bm \xi})$, where
${{\bm   \xi}  = (\xi_x,\, \xi_y)}$ is a two--coordinate on the screen,
see Fig.~\ref{fig:scintillation_screen}. For definiteness, we assume
that the scintillations occur between the lens and the observer at
distances $d_{lS}$ and $d_{So}$; we will comment on the
other choice below. Physically, the scintillations may happen in the
FRB host galaxy and / or the Milky Way. 

Since the scintillation phase $\varphi_S$ is caused by refraction, it
is given by Eq.~(\ref{eq:17}). But now $\delta n_e$ is a random
component of the electron overdensity. It is customary to assume that
this component has a homogeneous and isotropic Kolmogorov
turbulent spectrum,
see~\cite{1985ApJ...288..221C},~\cite{Rickett1990}, \cite{Narayan}, 
\cite{Lorimer-Kramer}, \cite{Woan},  \cite{Katz:2019qug}:
\begin{multline}
  \label{eq:5}
  \langle \delta n_e (\bm{X})\,  \delta n_e(\bm{X}')\rangle \\= C_n^2
  \int_{\kappa_{\mathrm{out}}}^{\kappa_{in}} d^3 
  \bm{\kappa} \; |\bm{\kappa}|^{-11/3} \;
  \mathrm{e}^{i\bm{\kappa}(\bm{X}' - \bm{X})}\;,
\end{multline}
where $\bm{X}$ and $\bm{X}'$ represent the three--dimensional space
coordinates and $\kappa_{\mathrm{in}} \gg \kappa_{\mathrm{out}}$~---
the cutoff scales for turbulence. Angular brackets in Eq.~(\ref{eq:5})
average over   
realizations of the turbulent ensemble e.g.\ volumes within the
galaxy. Using Eq.~(\ref{eq:17}), one can turn Eq.~(\ref{eq:5}) into
a correlator of two $\varphi_S$'s~\citep{Rickett1990, Katz:2019qug},
\begin{equation}
  \label{eq:10}
  S_\nu(\bm{\xi} - \bm{\xi}') \equiv \langle \left[\varphi_S(\bm{\xi}) -
  \varphi_S(\bm{\xi}')\right]^2 \rangle = \frac{|\bm{\xi} -
  \bm{\xi}'|^{5/3}}{r_{\mathrm{diff}}^{5/3}}\;, 
\end{equation}
where we introduced the diffractive length scale
\begin{multline}
  \label{eq:9}
  r_{\mathrm{diff}} = 3.63\cdot 10^{10} \, \mbox{cm}\,
  \left(\frac{C_n^2 \cdot L}{ 10^{-4} \, \mbox{m}^{-20/3} 
  \cdot \mbox{kpc}} \right)^{-3/5} \\\times \left(\frac{\nu}{6\,
    \mbox{GHz}}\right)^{6/5}\;, 
\end{multline}
and thickness of the scintillation screen $L\sim \mbox{kpc}$.
Equation (\ref{eq:10}) implies that the rays crossing the screen at
distance $r_{\mathrm{diff}}$ receive relative random phases of order
1. This makes them incoherent at $|\bm{\xi} - \bm{\xi}'| \gtrsim
r_{\mathrm{diff}}$. Technically, it will
be important for us that the correlator Eq.~(\ref{eq:10}) is translationally
invariant, i.e.\ depends on $|\bm{\xi} - \bm{\xi}'|$, and proportional
to $\nu^{-2}$,  cf.\ Eq.~(\ref{eq:17}).

In this Appendix we describe the scintillations statistically
i.e.\ compute the mean FRB spectra and their correlation functions. Recall
that in the main text we average the experimental data over the frequency
relying on the fact that they become statistically independent if $\nu$
is shifted by $\nu_d$ --- the decorrelation bandwidth. This approach
is applicable for narrow--band scintillations of
Sec.~\ref{sec:scintillations} that have small $\nu_d$ compared to the
total FRB bandwidth  $\nu_{\mathrm{bw}} \sim \mbox{GHz}$. The same
description is at best qualitative, however, in the case of
wide--band  scintillations considered in
Sec.~\ref{sec:progenitor-spectrum}. Below we heavily rely on the
expansion  in~${\nu_d/\nu \ll 1}$.  

In the thin--screen approach of Fig.~\ref{fig:scintillation_screen} the
radio ray $p\bm{x\xi}o$ consists of three straight parts. Its
geometric phase shift can be written as
\begin{equation}
  \label{eq:30}
  \phi = \phi_l(\bm{x}) + \phi_S(\bm{x},\, \bm{\xi})\;,
\end{equation}
where $\phi_l$ is the same lens shift Eq.~(\ref{eq:18}) as before and
\begin{equation}
  \label{eq:31}
  \phi_S = 2\pi \nu(r_{\bm{x}\bm{\xi}} + r_{\bm{\xi}o} - r_{\bm{x}o})
  \approx (\bm{\xi} - \tilde{\bm{\xi}})^2/2r_{F,\, S}^2 
\end{equation}
accounts for the additional turn at the point $\bm{\xi}$ of the
scintillation screen. We introduced the coordinate
\begin{equation}
  \label{eq:32}
  \tilde{\bm{\xi}}(\bm{x}) = (d_{So}\bm{x} + d_{lS} \bm{x}_o)/d_{lo}
\end{equation}
corresponding to straight propagation between $\bm{x}$ and
$o$. Besides,
\begin{equation}
  \label{eq:56}
  r_{F,\, S} = (d_{lS}d_{So}/2\pi \nu
  d_{lo})^{1/2}
\end{equation}
is the Fresnel scale for scintillations.

To sum up, we consider radio wave which sequentially crosses the
lens and the scintillating medium acquiring the total phase $\Phi =
\phi_l +  \varphi_l + \phi_S + \varphi_S$. The Fresnel integral for
this wave has the form, cf.\ Eq.~(\ref{eq:19}), 
\begin{equation}
  \label{eq:33}
  f_\nu = f_p \int \frac{d^2 \bm{x}}{{2\pi i r_{F,\, l}^2}} \;\; \mathrm{e}^{i
  \phi_l(\bm{x}) + i\varphi_l(\bm{x})} \; {\cal A}_\nu(\bm{x})\;,
\end{equation}
where the new factor ${\cal A}_\nu$ in the integrand accounts for
scintillations,
\begin{equation}
  \label{eq:34}
  {\cal A}_{\nu} = \int \frac{d^2 \bm{\xi}}{2\pi i r_{F,\, S}^2} \;\;
  \mathrm{e}^{i \phi_S(\bm{x},\, \bm{\xi}) + i \varphi_S(\bm{\xi})}\;.
\end{equation}
Recall that $\varphi_S$ and hence $f_{\nu}$ are the random
quantities.

\subsection{Mean fluence}
\label{sec:mean-fluence}
The Fresnel integral for the averaged fluence ${\langle F \rangle = \langle
f_\nu f_\nu^*\rangle}$ runs over $\bm{x},\; \bm{\xi}$ and $\bm{x}',\;
\bm{\xi}'$ which come from the integrals~(\ref{eq:33}), (\ref{eq:34})
for $f_\nu$ and $f_\nu^*$, respectively. In this expression, the
statistical mean $\langle \cdot \rangle$ acts on the random phase
${\exp[i\varphi_S(\bm{\xi}) -    i\varphi_S(\bm{\xi}')]}$ in the
integrand. Recall that we consider 
weakly interacting turbulent plasma with $\varphi_S \propto
\delta n_e$ behaving as a Gaussian random quantity. Any correlator of
such  quantity can be computed in terms of the two--point
function~(\ref{eq:10}). In particular,
\begin{equation}
  \label{eq:35}
  \langle \mathrm{e}^{i\varphi_S(\bm{\xi}) - i\varphi_S(\bm{\xi}')}\rangle =
  \mathrm{e}^{-S_\nu (\bm{\xi} - \bm{\xi}')/2}\;.
\end{equation}
Thus, the scintillation factor in the integrand of $\langle F\rangle$
equals,
\begin{equation}
  \notag
  \langle {\cal A}_{\nu} {\cal A}_\nu'^*\rangle  = \int \frac{d^2
    \bm{\xi} \, d^2\bm{\xi}'}{4\pi^2 r_{F,\, S}^4}  \;\;
  \mathrm{e}^{i\phi_S(\bm{x},\, \bm{\xi}) -
  i \phi_S(\bm{x}',\, \bm{\xi}') - S_\nu (\bm{\xi} - \bm{\xi}')/2}\;,
\end{equation}
where ${\cal A}_\nu' \equiv {\cal A}_\nu(\bm{x}')$. From
Eq.~(\ref{eq:31}) one learns that 
$\bm{\xi} + \bm{\xi}'$ enters linearly the exponent. As a consequence,
the integral over this combination produces ${\delta^{(2)} [\bm{\xi} -
  \bm{\xi}' - \tilde{\bm{\xi}}(\bm{x}) + \tilde{\bm \xi}(\bm 
x')]}$, and we obtain,
\begin{equation}
  \label{eq:37}
  \langle {\cal A}_{\nu} {\cal A}_\nu'^*\rangle = \exp\left(-
  \frac{|\bm{x} - \bm{x}'|^{5/3}}{2\tilde{r}_{\mathrm{diff}}^{5/3}}\right)\;, 
\end{equation}
where a projection
\begin{equation}
  \label{eq:57}
  \tilde{r}_{\mathrm{diff}} =
  d_{lo} r_{\mathrm{diff}}/d_{So}
\end{equation}
of the diffractive scale onto the lens screen was introduced.

We conclude that the net effect of scintillations  is to ruin
coherence in $\langle F\rangle$ i.e.\ suppress contributions of radio
paths $\bm{x}$ and $\bm{x}'$  if the distance between them
exceeds $\tilde{r}_{\mathrm{diff}}$. Using Eqs.~(\ref{eq:33}),
(\ref{eq:37}), we write the mean fluence as  
\begin{equation}
  \label{eq:38}
  \langle F \rangle = F_p\int \frac{d\bm{x}  d\bm{x}'}{4\pi^2 r_{F,\,
      l}^4} \, \mathrm{e}^{i\Phi_l 
    - i\Phi_l'  - \frac12 (|\bm{x}
    - \bm{x}'| / \tilde{r}_{\mathrm{diff}})^{5/3}}\;,
\end{equation}
where $F_p \equiv |f_p|^2$, $\Phi_l \equiv \phi_l(\bm{x}) +
\varphi_l(\bm{x})$, and $\Phi_l' \equiv \Phi_l(\bm{x}')$. Note
that the lens lurking in the FRB host galaxy is almost insensitive to
the scintillations in the Milky Way:
$\tilde{r}_{\mathrm{diff}}/r_{\mathrm{diff}} = d_{lo}/d_{So} \sim 
10^{6}$. 

If the scintillations occur between the source and the lens, the
integral (\ref{eq:38}) is still valid but the projected diffractive
scale equals $\tilde{r}_{\mathrm{diff}} =
d_{pl}r_{\mathrm{diff}}/d_{pS}$. Then the scintillations in
the FRB host galaxy are geometrically suppressed if the lens is near
us. 

Now, recall that we consider large--size lenses in the limit of
geometric optics $\Phi_l \gg 1$, cf.\ Appendix~\ref{sec:lenses}. At
the same time,  we are interested in detectably large  contributions
i.e.\ in the situation when the scintillation factor is not
too small i.e.\ ${|\bm{x} -\bm{x}'| / \tilde{r}_{\mathrm{diff}}
  \lesssim
  \mbox{few}}$. In this case the integral is dominated by the same
stationary points\footnote{Strongly suppressed contributions can be
  computed by finding complex extrema $(\bm{x},\; \bm{x}')$ of the
  full exponent in Eq.~(\ref{eq:38}).} $\bm{x} = \bm{x}_j$ and
$\bm{x}' = \bm{x}_{j'}$~--- the paths of radio rays~--- as in the case
without the scintillations. We obtain, 
\begin{multline}
  \label{eq:39}
  \langle F \rangle = F_p\sum\limits_{jj'} (G_j G_{j'})^{1/2} \,
  \mathrm{e}^{i\Phi_l (\bm{x}_j) - i\Phi_l(\bm{x}_{j'})} \\ \times
  \mathrm{e}^{ - \frac12 (|\bm{x}_j 
    - \bm{x}_{j'}| / \tilde{r}_{\mathrm{diff}})^{5/3}}
\end{multline}
where $\bm{x}_j$ solve the lens equation~(\ref{eq:20}) and the
gain factors $G_j$ are given by the same determinant as before. The
effect of the scintillations is represented by the exponent in the
second line of Eq.~(\ref{eq:39}).

In the simplified case of two radio paths one obtains the analog of
Eq.~(\ref{eq:24}),
\begin{equation}
  \label{eq:40}
  \langle F \rangle = \bar{F}_{50} \Big[ 1 + A_{\mathrm{osc}} \; 
    \mathrm{e}^{ - \frac12 (|\bm{x}_1
    - \bm{x}_2| / \tilde{r}_{\mathrm{diff}})^{5/3}}\cos(\Phi_1 -
    \Phi_2)\Big]\;,
\end{equation}
where the only new factor is a scintillating exponent suppressing
interference between the trajectories with $|\bm{x}_1 - \bm{x}_2|
\gtrsim \tilde{r}_{\mathrm{diff}}$.  Note that it is hard to compare
Eq.~(\ref{eq:40}) with experiment. We already explained that the
only\footnote{Another option would include averaging over many burst
  spectra. But we cannot do that: we have only 18 bursts which are
  unevenly distributed in time and have essentially different
  wide--band structure.} practical way to perform the statistical
average is to smooth over frequency. But~--- alas~--- the same
procedure kills the oscillating interference terms. Thus, one either
has to consider sophisticated statistical methods  like
Kolmogorov--Smirnov--Kuiper test~\citep{NR} or analyze more involved
spectral correlators.

\subsection{Autocorrelation function}
\label{sec:autoc-funct}
Now, consider the mean product of fluences at nearby frequencies $\nu$ and
$\nu_1 = \nu+ \Delta \nu$,
\begin{equation}
  \label{eq:41}
  \langle F(\nu) F(\nu_1) \rangle = \langle f_\nu f_\nu^*
  f_{\nu_1} f_{\nu_1}^* \rangle\;.
\end{equation}
Using Eqs.~(\ref{eq:33}), (\ref{eq:34}) for the amplitudes, we write
it in the form of a Fresnel integral over $(\bm{x},\;
\bm{\xi})$, $(\bm{x}',\; \bm{\xi}')$, $(\bm{x}_1,\; \bm{\xi}_1)$, and
$(\bm{x}_1',\; \bm{\xi}_1')$. Here and below we mark all quantities
related to the last three amplitudes $f_{\nu}^*$, $f_{\nu_1}$,
and $f_{\nu_1}^*$ by prime, 1, and 1-prime, respectively.

\begin{figure}
  \centerline{\includegraphics{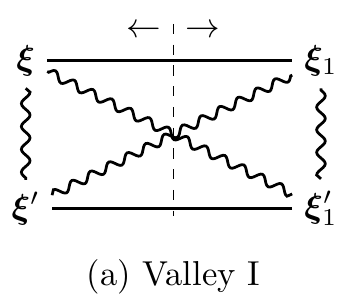} \hspace{5mm}
    \includegraphics{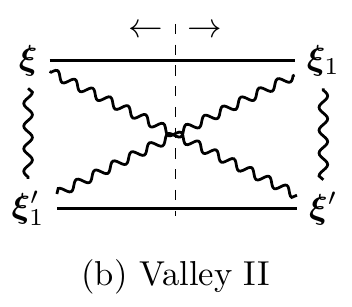}}
  \caption{Two graphical representations of
    Eq.~(\ref{eq:43}).
  \label{fig:diag4}}
\end{figure}
The integrand of the above four--wise Fresnel integral involves
a statistical average of four random phases 
\begin{equation}
  \label{eq:42}
  \mathrm{e}^{-S_4/2} \equiv \langle \mathrm{e}^{i\varphi_S -
    i\varphi_S' + i\varphi_{S1} - i\varphi_{S1}'} \rangle \;,
\end{equation}
where $\varphi_{S1}' \equiv \varphi_S(\bm{\xi}_1')$, etc. Like before,
this correlator can be computed by recalling that $\varphi_S$ is a 
Gaussian random variable and exploiting Eq.~(\ref{eq:10}),
\begin{align}
\label{eq:43}
  S_4 =&  \langle \left(\varphi_S - \varphi_S'
    + \varphi_{S1} - \varphi_{S1}' \right)^2
  \rangle\\ \notag
   =  &S_\nu(\bm{\xi} - \bm{\xi}') + \frac{\nu^2}{\nu_1^2} \;S_{\nu}
  (\bm{\xi}_1 - \bm{\xi}_1')  + \frac{\nu}{\nu_1} 
  S_\nu(\bm{\xi}_1'- \bm{\xi}) \\ &+\frac{\nu}{\nu_1} \Big[
    S_\nu(\bm{\xi}_1 - \bm{\xi}') - 
    S_\nu(\bm{\xi}_1 - \bm{\xi}) - S_\nu(\bm{\xi}_1' -
    \bm{\xi}')\Big]\;, \notag
\end{align}
where we explicitly used the fact that $\varphi_S \propto \nu^{-1}$,
cf.\ Eq.~(\ref{eq:17}). Expression~(\ref{eq:43})
looks complicated. That is why we visualize it in Fig.~\ref{fig:diag4} by
drawing every $S_{\nu}(\bm{\xi} - \bm{\xi}')$ with an ``attractive''
spring between $\bm{\xi}$ and~$\bm{\xi}'$, and~$(-S_\nu)$~--- with a
``repulsive'' solid line.

The exponent~(\ref{eq:42}) cuts off the $\bm{\xi}$ integrations in
regions with large $S_4$. Generically, this happens if the distances
between $\bm{\xi}$'s exceed $r_{\mathrm{diff}}$. However, there are
two valleys~--- large integration regions with small $S_4$. First, one
can keep $(\bm{\xi},\; \bm{\xi}')$ and $(\bm{\xi}_1,\, \bm{\xi}_1')$
in tight pairs increasing the distance $\bm{R} = \bm{\xi}_1 -
\bm{\xi}$ between them, see ``Valley~I'' in Fig.~\ref{fig:diag4}a. In
this case
\begin{equation}
  \label{eq:44}
  S_{4}\big|_{\mathrm{valley}\;  \mathrm{I}} \approx S_\nu (\bm{\xi} -
    \bm{\xi}') + S_{\nu_1}(\bm{\xi}_1 - \bm{\xi}_1')\;,
\end{equation}
with corrections of order $(r_{\mathrm{diff}}/|\bm{R}|)^{1/3}$.
Second, we can move apart the pairs $(\bm{\xi},\, \bm{\xi}_1')$ and
$(\bm{\xi}_1, \, \bm{\xi}')$ in Fig.~\ref{fig:diag4}b. This gives
\begin{multline}
  \label{eq:45}
  S_4\big|_{\mathrm{valley}\; \mathrm{II}} \approx (\Delta \nu/\nu)^2\,
  S_\nu(\bm{\xi}_1 - \bm{\xi}) + S_\nu(\bm{\xi} - \bm{\xi}_1') \\+
  S_\nu(\bm{\xi}_1 - \bm{\xi}')\;,
\end{multline}
with $\Delta \nu \equiv \nu_1 - \nu$ and similar corrections as
before. Soon we will see that\footnote{In a specific kinematic regime
  $|\tilde{\bm{x}} -\tilde{\bm{x}}_1 | \sim (\nu/\Delta \nu)^{6/5} \,
  r_{\mathrm{diff}}$ one correction in Valley II behaves as
  $(\nu_d/\nu)^{1/5} \sim 22\%$. However, this regime is irrelevant
  for the discussion in the main text.}
$(r_{\mathrm{diff}}/R)^{1/3} \sim (\nu_d/\nu)^{1/3} \sim 8\%$,  where
the numerical estimate is performed for narrow--band scintillations of 
Sec.~\ref{sec:scintillations}.

Now, consider the scintillation factor ${{\cal A}_4 \equiv \langle
{\cal A}_\nu {\cal A}_{\nu}'^*{\cal A}_{\nu_1}{\cal
  A}_{\nu_1}'^*\rangle}$ in the Fresnel integral for Eq.~(\ref{eq:41}). It
involves four $\bm{\xi}$--integrals,
\begin{equation}
  \label{eq:47}
  {\cal A}_4 = \int \frac{d^2\{\bm{\xi} \bm{\xi'} \bm{\xi}_1
      \bm{\xi}_1'\}}{(2\pi r_{F,\, S}^2)^4} \;\; \mathrm{e}^{i \phi_S
    - i\phi_S' + i \phi_{S1} - i \phi_{S1}'-S_4/2}\;,
\end{equation}
see Eq.~(\ref{eq:34}). Valleys I and II with paired $\bm{\xi}$'s give
major contributions into~${\cal A}_4$. From the technical viewpoint,
these valleys appear because the four scintillation factors correlate
in pairs and  their average product almost equals the sum of $\langle
{\cal A}_\nu  
{\cal A}_{\nu}'^*\rangle\langle  {\cal   A}_{\nu_1} {\cal
  A}_{\nu_1}'^*\rangle$ and $\langle {\cal A}_\nu {\cal
  A}_{\nu_1}'^*\rangle\langle  {\cal A}_{\nu_1} {\cal
  A}_{\nu}'^*\rangle$, with two--point correlators decaying
exponentially at far--away $\bm{\xi}$'s.  

Let us start with the valley I, Eq.~(\ref{eq:44}). In this case the
averaged scintillation phase $\exp(-S_4/2)$ depends separately  on
$(\bm{\xi},\, \bm{\xi}')$ and $(\bm{\xi}_1,\,
\bm{\xi}_1')$. Integrating over these pairs in the same way as in
Sec.~\ref{sec:mean-fluence}, we find,
\begin{equation}
  \label{eq:50}
  {\cal A}_4\big|_{\mathrm{valley}\; \mathrm{I}} \approx \mathrm{e}^{-
  \frac12 (|\bm{x} - \bm{x}'|/\tilde{r}_{\mathrm{diff}})^{5/3} - \frac12 
    (|\bm{x}_1 - \bm{x}_1'| / \tilde{r}_{\mathrm{diff}})^{5/3}}\!,
\end{equation}
where $\tilde{r}_{\mathrm{diff}}$ is the projected diffraction
scale Eq.~(\ref{eq:57}).  Due to Eq.~(\ref{eq:50}),  the entire
correlator Eq.~(\ref{eq:41}) factorizes along valley~I into 
\begin{equation}  
  \label{eq:46}
  \langle F(\nu)F(\nu_1) \rangle\big|_{\mathrm{valley}\; \mathrm{I}} \approx
  \langle F(\nu) \rangle \langle F(\nu_1)\rangle\;,
\end{equation}
cf.\ Eq.~(\ref{eq:38}). 

The valley II is a bit different. The exponent $S_4$
in Eq.~(\ref{eq:45}) weakly depends on the variable ${\bm{R} \equiv
\bm{\xi}_1 - \bm{\xi}}$ along this valley because the random
phases do not completely compensate in the products  ${\cal A}_{\nu}{\cal
  A}_{\nu_1}'^*$ and  ${\cal  A}_{\nu_1}{\cal A}_{\nu}'^*$  at
$\nu_1 \ne \nu$. Nevertheless, we will see that the
$\bm{R}$--dependence factorizes: the factor $(\Delta \nu/\nu)^2 \ll
1$ in front of the first term in Eq.~(\ref{eq:45}) makes it
insensitive to the small displacements $\delta  \bm{\xi} =  \bm{\xi} -
\bm{\xi}_1'$ and $\delta \bm{\xi}_1 = \bm{\xi}_1 - \bm{\xi}'$
transverse to the valley. Indeed, notice that due to the shift symmetry
the center--of--mass variable $\bm{\xi} + \bm{\xi}' + \bm{\xi}_1 +
\bm{\xi}_1'$ enters linearly in the exponent of Eq.~(\ref{eq:47}) and
therefore gives  $$
\delta^{(2)} \left[\bm{R} - \tilde{\bm{R}} + \frac{\nu}{\Delta \nu}
  \,(\delta \bm{\xi} + \delta \bm{\xi}_1 - \delta\tilde{\bm{\xi}} -
  \delta\tilde{\bm{\xi}}_1)\right] \;,
$$
where we introduced $\tilde{\bm{R}} = \tilde{\bm{\xi}}_1 -
\tilde{\bm{\xi}}$, $\delta \tilde{\bm{\xi}}  =  \tilde{\bm{\xi}} -
\tilde{\bm{\xi}}_1'$, etc., like in the case without the tildes. This
fixes the value of~$\bm{R}$  
leaving only the integrals over $\delta \bm{\xi}$ and $\delta
\bm{\xi}_1$ in Eq.~(\ref{eq:47}). Notably, $\delta \bm{\xi},\; \delta 
\bm{\xi}_1 \sim r_{\mathrm{diff}}$, and thus ${|\bm{R}|/r_{\mathrm{diff}}
\sim \nu/\Delta \nu \sim \nu/\nu_d}$, like was announced previously.
Moreover, we indeed can use $\bm{R} \approx \tilde{\bm{R}} \equiv (\bm{x}_1 -
\bm{x}) d_{So}/d_{lo}$ in the first term of Eq.~(\ref{eq:47})
ignoring corrections because this term is already suppressed by the
factor $(\Delta \nu/\nu)^2$.

Once this is done, dependences on  $\delta\bm{\xi}$ and $\delta
\bm{\xi}_1$ factorize and we arrive to  
\begin{multline}
  \label{eq:48}
        {\cal A}_4\big|_{\mathrm{valley}\; \mathrm{II}}
        =  \exp\left\{-\frac{\Delta \nu^2}{2\nu^2}\, \frac{|\bm{x}
          - \bm{x}_1|^{5/3}}{\tilde{r}_{\mathrm{diff}}^{5/3}}\right\}
        \\ \times
        {\cal J}_{\Delta \nu}\left(\frac{|\bm{x} -
          \bm{x}_1'|}{\tilde{r}_{\mathrm{diff}}}\right)\;  {\cal
          J}_{\Delta \nu}^*\left(\frac{|\bm{x}_1 -
          \bm{x}'|}{\tilde{r}_{\mathrm{diff}}}\right) \;,
\end{multline}
where the functions ${\cal J}_{\Delta \nu}(\rho)$ and ${\cal J}^*_{\Delta
  \nu}(\rho_1)$ include integrals over $\delta \bm{\xi}$ and $\delta
\bm{\xi}_1$. The latter can be written in the simplified form\footnote{We
  integrated over orientations of   $\delta 
  \bm{\xi}$ and introduced ${\zeta =   |\delta
  \bm{\xi}|^2 / (w r_{\mathrm{diff}}^2)}$.}
\begin{multline}
  {\cal J}_{\Delta \nu}(\rho)  = -i\,  \mathrm{e}^{i\rho^2
    /w} \\ \times \int_{0}^{\infty} d\zeta \; \mathrm{e}^{i\zeta  -
    \frac12 (w\zeta)^{5/6}}   J_0
  \left(2 \rho\sqrt{\zeta/w}\right)\;.
\end{multline}
We introduced the rescaled frequency lag $w \equiv 2\Delta\nu/\nu_d$,
decorrelation bandwidth ${\nu_d \equiv \nu\,(r_{\mathrm{diff}}/r_{F,\,
    S})^2}$, the argument $\rho$ is of ${\cal J}_{\Delta \nu}$
measuring  distances between $\bm{x}$'s in units of
$\tilde{r}_{\mathrm{diff}}$, and the Bessel function $J_0$. Note that
the ${\cal J}$--factors in 
Eq.~(\ref{eq:48}) do the same job as the exponent in
Eq.~(\ref{eq:50}): they force $|\bm{x} 
- \bm{x}_1'|$ and  $|\bm{x}_1 - \bm{x}'|$ to be smaller than
$\tilde{r}_{\mathrm{diff}}$. Indeed, ${\cal J}_{\Delta \nu}(\rho) \to  \exp\{
  -\frac12 \rho^{5/3}\}$ as  $\rho \to +\infty$.  At $\rho \lesssim
w$ this function remains unsuppressed and essentially depends on the
frequency lag $w\propto \Delta \nu$.

We finally consider the integrals over the lensed radio paths
$\bm{x}$, $\bm{x}'$, $\bm{x}_1$, and $\bm{x}_1'$ in the Fresnel
representation of the correlator Eq.~(\ref{eq:41}),
cf.\ Eq.~(\ref{eq:33}). The scintillations supply the factor in the
integrand, 
\begin{equation} 
  \label{eq:52}
  {\cal A}_4(\bm{x},\, \bm{x}',\, \bm{x}_1,\, \bm{x}_1') =  {\cal
    A}_4\big|_{\mathrm{valley}\; \mathrm{I}} + {\cal
    A}_4\big|_{\mathrm{valley}\; \mathrm{II}}\;,
\end{equation}
given by the sum of Eqs.~\eqref{eq:50} and~\eqref{eq:48}.

Like in the previous Sections, we consider only large--size lenses in
the limit of geometric optics. In this case the phase shift is large,
${\Phi_l \sim |\bm{x} - \tilde{\bm{x}}|^2/   r_{F,\,   l}^2 \gg 1}$, and
the Fresnel integral is dominated by the set of distinguished rays $\{
\bm{x}_j \}$ satisfying the lens equation~\eqref{eq:20}. Notably, one
can roughly estimate the typical lens phase via Eq.~(\ref{eq:23})
obtaining $\Phi_l \sim 2\pi \nu /T_\nu $, where $T_\nu$ is a period of
spectral oscillations. Thus, the geometric optics is indeed valid in
the case $T_\nu\ll \nu$ considered in the main text. The
saddle--point integration gives,
\begin{multline}
  \label{eq:49}
  \langle F(\nu) F(\nu_1)\rangle= F_p^2\sum\limits_{j j' j_1 j_1'} (G_j G_{j'}
  G_{j_1} G_{j_1'})^{1/2} \\ \times  \mathrm{e}^{i \Phi_l(\bm{x}_j) -
    i\Phi_l(\bm{x}_{j'}) + i\Phi_l(\bm{x}_{j_1}) -
    i\Phi_l(\bm{x}_{j_1'})} \cdot {\cal A}_{4}\;,
\end{multline}
where the sum runs four--wise over the radio paths $\bm{x}_j$, the
scintillation factor ${\cal A}_4$ depends on the four of them, we
introduced the lens phases $\Phi_l(\bm{x}_j)$ and the gain  factors
$G_j$, see Sec.~\ref{sec:lenses}. Note that ${\cal A}_4$ selectively
suppresses the interference terms with far--away $\bm{x}$'s. 

Despite the complex form, Eq.~(\ref{eq:49}) is easy to use. Indeed, it
involves the same radio rays as in the previous Section, whereas the 
suppression factor ${\cal A}_4$ is explicitly given by Eqs.~\eqref{eq:52},
\eqref{eq:50}, and \eqref{eq:48}. Now, we continue with examples. 

\subsection{ACF for scintillations only}
\label{sec:scint-but-no}
Suppose first that the lens is absent, $\varphi_l = 0$. In this case
there exists only one radio ray $\bm{x} = \tilde{\bm{x}}$
corresponding to straight propagation between the source and
the scintillation screen. We obtain $\Phi_l=0$ and $G
=1$. Expression~\eqref{eq:49} then reduces to the scintillation
factor~\eqref{eq:50}, \eqref{eq:48}, 
\begin{equation}
  \label{eq:51}
  \langle F(\nu) F(\nu + \Delta \nu)\rangle = F_p^2  +F_p^2 \left|h\left(
  2\Delta \nu /\nu_d\right)\right|^2 \;,
\end{equation}
where the decorrelation bandwidth $\nu_d(\nu)$ is given by
Eq.~(\ref{eq:54}), and we introduced the 
hat--like function
\begin{equation}
  \label{eq:53}
  h(w) \equiv {\cal J}_{\Delta \nu}(0) = -i \int_{0}^{+\infty} d\zeta
  \; \mathrm{e}^{i\zeta - \frac12 (w\zeta)^{5/6}} 
\end{equation}
shown in Fig.~\ref{fig:h} (solid line). One can explicitly check that
in the asymptotic regions $w\to 0,\; +\infty$ this function coincides
with the approximation Eq.~(\ref{eq:75}) used in
Sec.~\ref{sec:scintillations}. The latter correctly represents $h(w)$
even at finite $w$, see the dashed line in Fig.~\ref{fig:h}. 

Importantly, $|h(w)|$ falls off from 1 at $w=0$ to zero at large
frequency lags. As a consequence, the  correlator Eq.~(\ref{eq:51})
monotonically decreases from  $2F_{p}^2$ at $\Delta \nu=0$ to 
$F_p^2$ at large $\Delta \nu$ reflecting the fact that $F(\nu)$ and
$F(\nu_1)$ are not correlated at far--away 
frequencies.  Besides, ${|h|^2 = 1/2}$ at $\Delta \nu\approx 0.96 \,
\nu_d$. Thus, up to a factor 0.96 our definition of $\nu_d$
  coincides with the standard one.

Using $F_p\approx \bar{F}_{50}(\nu)$ and Eq.~(\ref{eq:25}), we obtain 
the autocorrelation function Eq.~\eqref{eq:55} from the main text. Note
that the argument $w = 2\Delta \nu/\nu_d$ of $h$ sharply depends on
the frequency: $\nu_d \propto \nu^{22/5}$, see
Eq.~(\ref{eq:9}). When comparing with the experiment,
one has to take this dependence into account, like we do in
Sec.~\ref{sec:scintillations}. 

\subsection{ACF for scintillations $+$ lensing}
\label{sec:two-radio-paths}
Now, suppose  the lens produces two coherent images of the source
$\bm{x}_1$ and $\bm{x}_2$ with phases $\Phi_1$, $\Phi_2$ and gain 
factors $G_1$, $G_2$. After passing the lens these rays go through the
scintillating plasma. One can imagine two opposite situations.

First, the rays may cross the plasma at close distances
$|\tilde{\bm{\xi}}_2 - \tilde{\bm{\xi}}_1| \ll r_{\mathrm{diff}}$
i.e.\ $|\bm{x}_2 - \bm{x}_1| \ll \tilde{r}_{\mathrm{diff}}$. The 
scintillations do not decohere these paths and ${\cal A}_4$
can be evaluated at zero transverse shifts in Eqs.~(\ref{eq:50}),
(\ref{eq:48}). Factorizing ${\cal   A}_4 = 1+|h|^2$,  one obtains
Eq.~\eqref{eq:59} from the main text~--- a generalization of
Eq.~(\ref{eq:26}) to the model with scintillations. Note that all
terms oscillating with frequency disappear from Eq.~\eqref{eq:59}  due
to the overall integral over~$\nu$.

It is worth recalling that the inter-path distance is related to the
frequency period $T_\nu$  of interference oscillations by
Eq.~\eqref{eq:23}: $|\bm{x}_2 - \bm{x}_1|^2 \sim 2\pi \nu r_{F,\, l}^2
/T_\nu$. With this formula, one can rewrite the coherent--paths condition
${|\bm{x}_2 - \bm{x}_1| \ll \tilde{r}_{\mathrm{diff}}}$ as
\begin{equation}
  \label{eq:58}
  \frac{r_{F,\, l}^2}{r_{F,\, S}^2} \ll \frac{\nu_d\, T_\nu}{2\pi\nu^2}
  \; \frac{d_{lo}^2}{d_{So}^2}\;,
\end{equation}
where we used Eqs.~\eqref{eq:57}, \eqref{eq:54} and assumed that
scintillations occur between the lens and the observer. The
  experimental values of Secs.~\ref{sec:scintillations},
\ref{sec:periodic-structure} give
$\nu_d /\nu \sim 10^{-3}$ and $T_\nu / \nu \sim  10^{-2}$.
The inequality (\ref{eq:58}) is then
satisfied and the scintillations are irrelevant, say, if they occur in our
galaxy and the lens hides in the FRB host galaxy: $r_{F,\, l} \sim
r_{F,\, S}$ and $d_{lo}/d_{So} \sim 10^6$. The other possibilities
include Milky Way scintillations and the lens in the intergalactic space, or
scintillations in the FRB host galaxy and the lens outside of
it\footnote{In this last situation one replaces $d_{lo}/d_{So} \to
  d_{pl}/d_{pS}$ in Eq.~(\ref{eq:58}).}.

Second, at  $|\bm{x}_2 - \bm{x}_1| \gg \tilde{r}_{\mathrm{diff}}$
the decoherence of radio rays is relevant and scintillations kill the
majority of the interference terms. Expression~(\ref{eq:49}) takes the
form,\begin{widetext}
\begin{multline}
  \label{eq:60}
  \langle F(\nu) F(\nu_1) \rangle = \bar{F}_{50}^2 \Bigg[1 + |h|^2 + 
    \frac{A_{\mathrm{osc}}^2}{2}\, |h|^2
    \left( -1 + \cos(2\pi 
    \Delta \nu/T_\nu) \,\exp \left\{-\frac{\Delta
      \nu^2}{2\nu^2} \frac{|\bm{x}_2 -
      \bm{x}_1|^{5/3}}{\tilde{r}_{\mathrm{diff}}^{5/3}}\right\}\right)\Bigg]\;,
\end{multline}
\end{widetext}
where we introduced the same $\bar{F}_{50}$ and $A_{\mathrm{osc}}$ as
before, and $h \equiv h(2\Delta \nu /\nu_d)$. 
Notably, the oscillating term persists in this case due to imperfect
cancellation between the phases of $f_\nu$ at different
frequencies. However, the amplitude of oscillations decreases
with~$\Delta \nu$ becoming invisibly small at ${\Delta \nu \gg
  \nu_d}$ or $\Delta \nu \gg \nu \,\tilde{r}_{\mathrm{diff}}^{5/6} 
 / {|\bm{x}_2 - \bm{x}_1|^{5/6}}$. This means that
Eq.~(\ref{eq:60}) is relevant only at $T_{\nu} \ll \nu_d$, not  in
the case of Sec.~\ref{sec:scintillations}.

\bibliography{FRB}{} 
\bibliographystyle{aasjournal}

\end{document}